\documentclass[]{article}

\usepackage{cite}
\usepackage[a4paper]{geometry}
\usepackage[capposition=top]{floatrow}
\usepackage{amsmath}
\usepackage{amsfonts}
\usepackage{graphicx}
\usepackage{pzccal}
\usepackage{mathbbol}
\usepackage{natbib}
\newcommand{\edit}[1]{{\color{black}#1}}

\title{Synthesis estimators for \edit{transportability with} positivity violations \edit{by} a continuous covariate}
\author{Paul N Zivich\textsuperscript{1}, 
	Jessie K Edwards\textsuperscript{1}, 
	Bonnie E Shook-Sa\textsuperscript{2}, 
	Eric T Lofgren\textsuperscript{3}, \\
	Justin Lessler\textsuperscript{1,4,5},
	Stephen R Cole\textsuperscript{1}}
\date{%
	\small
	\textsuperscript{1}Department of Epidemiology, Gillings School of Global Public Health, University of North Carolina at Chapel Hill, Chapel Hill, NC\\%
	\textsuperscript{2}Department of Biostatistics, Gillings School of Global Public Health, University of North Carolina at Chapel Hill, Chapel Hill, NC\\%
	\textsuperscript{3}Paul G. Allen School for Global Health, Washington State University, Pullman, WA\\%
	\textsuperscript{4}Carolina Population Center, University of North Carolina at Chapel Hill, Chapel Hill, NC \\ %
	\textsuperscript{5}Department of Epidemiology, Johns Hopkins Bloomberg School of Public Health, Baltimore, MD\\[2ex]%
	\today
}

\begin{document}

\maketitle

\begin{abstract}
	Studies intended to estimate the effect of a treatment, like randomized trials, \edit{may not be sampled from} the desired target population. To correct for this \edit{discrepancy}, estimates can be transported to the target population. Methods for transporting between populations are often premised on a positivity assumption, such that all relevant covariate patterns in one population are also present in the other. However, eligibility criteria, particularly in the case of trials, can result in violations of positivity \edit{when transporting to external populations}. To address nonpositivity, a synthesis of statistical and mathematical models can be considered. This approach integrates multiple data sources (e.g. trials, observational, pharmacokinetic studies) to estimate treatment effects, leveraging mathematical models to handle positivity violations. This approach was previously demonstrated for positivity violations by a single binary covariate. Here, we extend the synthesis approach for positivity violations with a continuous covariate. For estimation, two novel augmented inverse probability weighting estimators are proposed. Both estimators are contrasted with other common approaches for addressing nonpositivity. Empirical performance is compared via Monte Carlo simulation. Finally, the competing approaches are illustrated with an example in the context of two-drug versus one-drug antiretroviral therapy on CD4 T cell counts among women with HIV.\\
	
\end{abstract}

\section{Introduction}

Descriptive or causal parameters are defined relative to a specific population \citep{maldonado_estimating_2002}, commonly referred to as the target population. However, data on outcomes \edit{may not be available for a} sample of the target population. For example, randomized trials often consist of a convenience sample of patients that would be eligible for the treatment, which can be thought of as a random sample of some (perhaps poorly defined) population \edit{other than the target population}. A problem occurs when the population \edit{giving rise to the trial data} differs in composition from the target population by important predictors of the outcome \citep{degtiar_review_2023}. \edit{More specifically, when the distribution of effect measure modifiers differs between the populations}, inference for the \edit{external} population may differ from the desired target population and produce misleading conclusions. To account for the incongruence between populations, one can integrate multiple data sources. In the case where data come from different populations not overlapping in time or space \edit{(i.e., population contributing information on the outcomes is external to the target population)}, one can \textit{transport} estimates to the data source designated to be a random sample of the target population \citep{degtiar_review_2023}. Much of the methods for transportability are premised under a conditional exchangeability assumption (i.e., independence of population membership and potential outcomes given a set of covariates) \citep{westreich_transportability_2017,dahabreh_generalizing_2019,bareinboim_causal_2016}. This exchangeability assumption comes with a structural positivity assumption, such that all pertinent patterns of covariates in the target population are also present in the \edit{external} population (but not necessarily vice-versa) \citep{westreich_transportability_2017,zivich_transportability_2023}. These identification assumptions can be suspect due to practical constraints, like eligibility criteria or differences in which (or how) variables were measured across data sources.

Work on relaxing the usual transportability identification assumptions has primarily focused on violations of the exchangeability assumption by unobserved covariates. Lesko et al. used a sensitivity analysis to compare how results changed under differing distributions of variables measured in a trial but not in a sample of their target population that were thought to be important for transporting \citep{lesko_effect_2016}. Nguyen et al. also considered the case where a variable was measured in the sample of the \edit{external} population but not the sample of the target population \citep{nguyen_sensitivity_2017}, and were further illustrated \edit{using} data from an HIV trial \citep{nguyen_sensitivity_2018}. In more recent work, Dahabreh et al. proposed sensitivity analyses using bias functions that do not require the variable to be measured in either sample \citep{dahabreh_sensitivity_2023}. Other related work has instead replaced the usual exchangeability assumption with a proximal causal inference analog \citep{nilsson_proxy_2023}. Given certain conditions on these proxy variables \citep{tchetgen_tchetgen_introduction_2020}, the transported parameter of interest can be identified despite the existence of a known but unmeasured outcome predictor that differs between populations.

Prior work on addressing positivity violations has primarily focused on positivity for the action (i.e., exposure, treatment, intervention) in direct studies of the target population \citep{zhu_core_2021}. Much of this work has approached such positivity violations by either revising the parameter of interest to ensure positivity \citep{petersen_diagnosing_2012,li_addressing_2019,kennedy_nonparametric_2019,naimi_incremental_2021}, modifying the estimated propensity scores \citep{cole_constructing_2008,sturmer_treatment_2010}, or using extrapolation from statistical models \citep{zhu_core_2021,nethery_estimating_2019,zhu_addressing_2023}. Less work has been done to address violations of the positivity assumption for transportability. Due to the asymmetry of positivity for transportability and the potential lack of outcome measurements in the target population sample, some of the previous methods for positivity violations are not immediately applicable. 

To address nonpositivity, we previously proposed a data integration approach that combines multiple individual-level data sources with external information through a synthesis of statistical and mathematical models \citep{zivich_transportability_2023}. The synthesis approach addresses key inadequacies of other common solutions to nonpositivity, like restricting the target population or restricting the covariate set. In that work, corresponding g-computation and inverse probability weighting (IPW) estimators were proposed. However, this prior work was limited to cases where the positivity assumption was violated with a single binary covariate. Here, the synthesis approach is extended to positivity violations with a continuous covariate. Additionally, two new augmented inverse probability weighting (AIPW) estimators are proposed. The first AIPW estimator is based on estimating the parameters of a marginal structural model (MSM) and the second is based on estimating the conditional average causal effect (CACE). As shown here, these estimators are generalizations of using a statistical model to extrapolate to the target population, where the mathematical model allows for integration of information external to the available data.

The organization of the paper is as follows. In Section 2, the data structure and the standard identification conditions for transportability are reviewed, with a particular focus on positivity violations in the context of transportability. The proposed synthesis method and other common approaches to address nonpositivity are reviewed in Section 3. In Section 4, a simulation study is presented to empirically assess the properties of the competing estimators. Section 5 details an illustrative example on the effect of two-drug ART versus one-drug ART on CD4 cell count at 20 weeks. Finally, Section 6 summarizes the key results and describes areas for future work.

\section{Identification}

Let $Y_i^a$ represent the potential outcome under action $a \in \{0,1\}$ for unit $i$, $A_i$ be the observed action, and $Y_i$ be the observed outcome. Let $V_i$ and $W_i$ be covariates measured at baseline. Here, $V_i$ consists of a continuous variable and $W_i$ is left arbitrary (i.e., a set of discrete and/or continuous variables). Finally, let $R_i$ indicate whether unit $i$ was in a sample of the target population ($R_i = 1$) or the \edit{external} population ($R_i = 0$). \edit{Here, one could imagine $R=0$ indicating membership in a historical randomized trial and $R=1$ indicate the general population eligible for $A$ in the present day, such that a unit is ineligible to be in both populations.} The parameter of interest, the average causal effect in the target population, can then be expressed as 
\[\psi = E[Y^1 | R=1] - E[Y^0 | R=1]\]
To proceed, $\psi$ is expressed in terms of the observables without placing constraints on the distributions, referred to as nonparametric identification. A commonly used set of assumptions for identification when the observed data consists of $(R_i = 1, V_i, W_i)$ for $n_1$ independent observations and $(R_i = 0, V_i, W_i, A_i, Y_i)$ for $n_0$ independent observations are reviewed next \citep{westreich_transportability_2017,dahabreh_extending_2020}. Let the observed data in general be denoted by $O_i$ hereafter.

The first identification condition, causal consistency \citep{cole_consistency_2009}, links the potential outcomes to the observed outcomes, and can be expressed as
\begin{equation}
	Y_i = Y_i^a \text{ if } a = A_i
	\label{Eq1}
\end{equation}
Let $f(\cdot)$ denote the probability density function. The next conditions are for the action in the \edit{external} population,
\begin{equation}
	\begin{split}
		E[Y^a \mid V=v,W=w,R=0] = & E[Y^a \mid A=a,V=v,W=w,R=0] \\ 
		&\text{ for } a\in\{0,1\} \text{ and } f_{V,W,R}(v,w,R=0) > 0
	\end{split}
	\label{Eq2}
\end{equation}
\begin{equation}
	\Pr(A=a \mid V=v, W=w, R=0) > \epsilon > 0 \text{ for } a\in\{0,1\} \text{ and } f_{V,W,R}(v,w,R=0) > 0
	\label{Eq3}
\end{equation}
which correspond to treatment exchangeability and treatment positivity, respectively \citep{hernan_estimating_2006,zivich_positivity_2022}. When $A$ is marginally randomized and $V,W$ are measured at baseline (i.e., a randomized trial), these conditions can reasonably be assumed to hold by design. To transport between populations, the following assumptions are considered
\begin{equation}
	\begin{split}
		E[Y^a \mid V=v,W=w,R=1] = & E[Y^a \mid V=v,W=w,R=0] \\ 
		&\text{ where } f_{V,W,R}(v,w,R=1) > 0
	\end{split}
	\label{Eq4}
\end{equation}
\begin{equation}
	\Pr(R=0 \mid V=v, W=w) > \epsilon > 0 \text{ where } f_{V,W,R}(v,w,R=1) > 0
	\label{Eq5}
\end{equation}
where (\ref{Eq4}) is conditional sampling exchangeability and (\ref{Eq5}) is sampling positivity \citep{westreich_transportability_2017}. Under these conditions, we can re-express the parameter of interest as
\begin{equation*}
	\begin{split}
		E[Y^a \mid R=1] = & E[E[Y^a \mid V,W,R=1] \mid R=1] = E[E[Y^a \mid V,W,R=0] \mid R=1]\\
		& = E[E[Y^a \mid A,V,W,R=0] \mid R=1] = E[E[Y \mid A,V,W,R=0] \mid R=1]
	\end{split}
\end{equation*}
which follows from iterated expectations, (\ref{Eq4}) and (\ref{Eq5}), (\ref{Eq2}) and (\ref{Eq3}), and (\ref{Eq1}), respectively. Therefore $\psi$ is identified, i.e., expressible in terms of the observed data.

\subsection{Positivity Violations with Continuous Covariates}

The previous identification proof relied on sampling positivity for $E[Y^a \mid V=v,W=w,R=0]$ to be well-defined. Equation (\ref{Eq5}) would be violated if the \edit{external} population was restricted to individuals below a threshold, $V_i < v_0$, or above a threshold, $V_i > v_0$, that was present in the target population. An example of this scenario would be a trial restricted to persons with HIV who have CD4 cell counts below (or above) a pre-specified level at baseline \citep{hammer_trial_1996,hammer_controlled_1997}, with inference being made to a target population of HIV infected persons without restricting by CD4. Regardless of the direction of the threshold, there is no information to inform the relationship between $Y^a$ and $V$ in the region(s) outside the support of the \edit{external} population. \edit{In practice, positivity violations by $V$ are most concerning for $\psi$ when $V$ is an effect modifier on the additive scale.}

\begin{figure}
	\centering
	\caption{Visualization of the issue presented by sample nonpositivity}
	\includegraphics[width=0.85\linewidth]{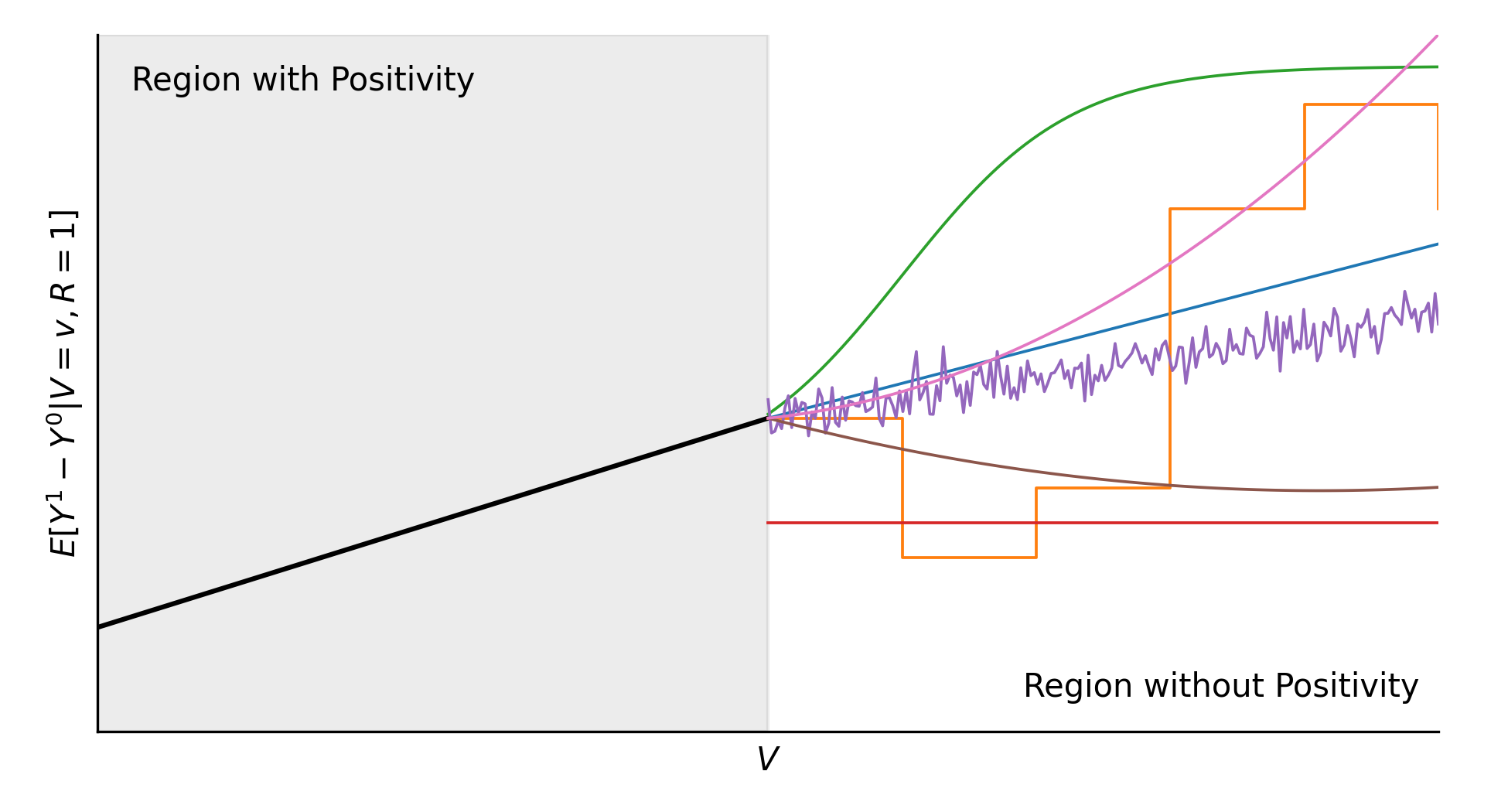}
	\floatfoot{The black line in the positive region is considered to be known. Colored lines in the nonpositive region are some examples of possible relationships for the nonpositive region. \edit{These alternatives cannot be ruled out by the observed data.}}
	\label{Fig1}
\end{figure}

To highlight this issue, consider the following structural sampling positivity violation; $\Pr(R = 0 \mid V^* = 1) = 0$ where $V^* = I(V > v_0)$. In this setting, Equation (\ref{Eq5}) does not hold for $V > v_0$. 
\edit{Even in the special case where the functional form of the causal effect by $V$ in the positive region is known, that model does not provide information for the nonpositive region. As illustrated in Figure \ref{Fig1}, a multitude of possible relationship by $V$ can exist across the nonpositive region.} Due to the violation of (\ref{Eq5}), none of these options can be ruled out without additional assumptions.

\section{Approaches to Nonpositivity}

To guide the subsequent discussion, consider the following re-expression of $\psi$ using the law of total expectation
\begin{equation}
	\psi = \psi_0 \Pr(V^* = 0 \mid R=1) + \psi_1 \Pr(V^* = 1 \mid R=1)
	\label{Eq6}
\end{equation}
where \edit{$V^* = 0$ indicates the region with positivity,} $\psi_0 = E[Y^1 - Y^0 | V^*=0, R=1]$, and $\psi_1 = E[Y^1 - Y^0 | V^*=1, R=1]$. Here, the sampling positivity violation means that $\psi_1$ is \textit{not} nonparametrically identifiable given Equations (\ref{Eq1})-(\ref{Eq5}). First, the proposed synthesis approach is \edit{described}, followed by popular alternatives.

\subsection{Synthesizing Statistical and Mathematical Models}

Equation (\ref{Eq6}) is used to motivate a combination, or synthesis, of statistical (e.g., regression, g-methods \citep{naimi_introduction_2017}) and mathematical (e.g., microsimulation \citep{krijkamp_microsimulation_2018,caglayan_microsimulation_2018}, mechanistic \citep{kirkeby_practical_2021,lessler_mechanistic_2016}) models. The underlying idea is to fit a statistical model for the regions with positivity, i.e. regions of the parameter space where an analog of (\ref{Eq5}) holds. Then a mathematical model is used to fill-in, or impute, outcomes across the nonpositive regions \edit{via external information}. In other words, one replaces $\psi_0$ with a statistical model and $\psi_1$ with a mathematical model.

First, notice that $\Pr(V^* = v^* \mid R=1)$ for $v^* \in \{0,1\}$ are already expressed in terms of the observables and thus are identified. Now consider modified versions of (\ref{Eq4}) and (\ref{Eq5}) that hold for $v$ where $V^*=0$,
\begin{equation}
	\begin{split}
		E[Y^a \mid V=v,W=w,R=1] = & E[Y^a \mid V=v,W=w,R=0] \\ 
		&\text{ where } f_{V,W,R}(v,w,V^*=0,R=1) > 0
	\end{split}
	\tag{4'}
	\label{Eq4a}
\end{equation}
\begin{equation}
	\Pr(R=0 \mid V=v, W=w) > \epsilon > 0 \text{ where } f_{V,W,R}(v,w, V^*=0,R=1) > 0
	\tag{5'}
	\label{Eq5a}
\end{equation}
Under these modified assumptions, $\psi_0 = E[E[Y \mid A=a,V,W, R=0] | V^*=0,R=1]$ (proof provided in Appendix 1).

As the \edit{external} population does not include anyone with $V^*=1$, $\psi_1$ is not nonparametrically identifiable with the available data. To progress, it is assumed that all actions considered, $a \in \{0,1\}$, are well-defined, or unambiguous, for everyone in the target population. If $\psi_1$ were known, then (\ref{Eq6}) could be directly combined with statistical models for the other components. Instead, a mathematical model is used to estimate $\psi$ by incorporating external information for the nonpositive region $\psi_1$. Let $\mathfrak{m}(a,V,W;\lambda)$ be a mathematical model for $Y^a$ defined by the parameters $\lambda$. For the mathematical model to stand-in for $\psi_1$, one could assume that $\mathfrak{m}(a,V,W;\lambda) = \psi_1$, but exact knowledge of $\psi_1$ is unlikely. So, this assumption is typically unreasonable. Instead, let $\mathfrak{M}_\lambda$ denote a set of distributions indexed by $\lambda$ for the functions of the potential outcomes, where $\mathfrak{m}(a,V,W;\lambda)$ is an element. Then one assumes that $\mathfrak{M}_\lambda$ contains the true conditional distribution of the potential outcomes for the nonpositive region, which can be expressed as
\begin{equation}
	\psi_1 \in \mathfrak{M}_\lambda.
	\label{Eq7}
\end{equation}
This assumption indicates that the structure and parameter values chosen for the mathematical model apply to the target population, i.e., it connects the parameterization of a mathematical model to the target population. Put another way, this assumption states that the structure and variations in the parameters of the set of mathematical models are rich enough to capture $\psi_1$. 

Discussion of how mathematical models can be constructed is delayed until after the introduction of both synthesis estimators. Both estimators are based on predictions for the region of positivity, which are then shifted by the mathematical model for the regions of nonpositivity. The distinction between the estimators is the type of external information required. The first estimator uses external information on the conditional mean of the potential outcomes and the second estimator uses external information on the CACE.

\subsubsection{Synthesis Estimator based on a Marginal Structural Model}

The synthesis estimator based on a MSM is defined as
\[\widehat{\psi}_{MSM} = \frac{1}{n_1} \sum_{i=1}^{n} \left[\mathcal{F}_1 (O_i; \hat{\alpha}, \hat{\eta}, \nu) - \mathcal{F}_0 (O_i; \hat{\alpha}, \hat{\eta}, \nu) \right] I(R_i = 1) \] 
where 
\[\mathcal{F}_a (O_i; \hat{\alpha}, \hat{\eta}, \nu) = t\left\{ s(a, O_i;\hat{\alpha}, \hat{\eta}) + m(a, O_i; \nu) \right\}\]
with $t\{\cdot\}$ denoting a user-specified transformation function (e.g., identity for linear models, inverse-logit for logistic models), $s(\cdot)$ is a statistical MSM with the parameters $\alpha$ and nuisance parameters $\eta$, and $m(\cdot)$ is mathematical model with $\lambda = \nu$ corresponding to shifts in the MSM. Here, hats denote quantities estimated using $O_i$.

To help contextualize this estimator and its additional flexibility over estimators proposed in the context of positivity violations with a binary covariate \citep{zivich_transportability_2023}, consider the following MSM specification for the target population
\[E[Y^a \mid V^*,R=1] = \text{expit}(\alpha_0 + \alpha_1 a + \nu_0 V^* + \nu_1 a V^*)\]
where $\alpha_0$ is the mean of $Y^0$ among $V^*=0$, and $\alpha_1$ is the average causal effect for $V^* = 0$. Then, $\nu_0$ is the difference in the mean under $a=0$ from $\alpha_0$ for those with $V^*=1$ (i.e., $\alpha_0 + \nu_0$ is the mean for $a=0$ among $V^*=1$). Finally, $\nu_1$ is the change in the average causal effect for $V^* = 1$. For this specification, the statistical model would be used to estimate $\alpha$, and $\nu$ comes from the mathematical model. However, suppose $X_i \in W_i$ and that the \edit{available} external information for $\nu$ was all conditional on $X$. As the parameters of a logistic model are non-collapsible \citep{greenland_confounding_1999,gail_biased_1984}, the external information that is conditional on $X$ is misspecified for the given MSM (i.e., the mathematical parameters are incorrect for the assumed MSM specification). Instead, the MSM for the target population could be revised to
\[E[Y^a \mid V^*, X, R=1] = \text{expit}(\alpha_0 + \alpha_1 a + \alpha_2 X + \alpha_3 aX + \nu_0 V^* + \nu_1 a V^*)\]
where the corresponding $\nu$ are now conditional on $X$. Therefore, the external information available matches the specified model. Note that this MSM implicitly assumes that at least one of the following is true: the distribution of $W_i \setminus {X_i}$ \edit{(i.e., all covariates in $W$ not in $X$)} does not differ by $V^*$ or the effect of $A$ is homogeneous by $W_i \setminus {X_i}$ for $V^* = 1$ (i.e., $\nu_2 = 0$ for $\nu_2 a X V^*$) for the full MSM to be correctly specified. These assumptions could be weakened by further refining the specification of the MSM. The ability to modify the MSM specification is a useful feature for this estimator as one does not often have the luxury of determining what external information is available. 

While the parameters of the statistical MSM can be estimated a number of different ways \citep{snowden_implementation_2011,robins_marginal_2000}, a weighted regression AIPW estimator is used here \citep{bang_doubly_2005,robins_comment_2007,vansteelandt_invited_2011}, with the Snowden et al. method for estimation of the MSM parameters \citep{snowden_implementation_2011}. First, the overall weights
\[\pi(O) = \frac{1}{\Pr(A=a \mid V,W,R=0, V^* =0)} \times \frac{\Pr(R=1 \mid V,W,V^*=0)}{\Pr(R=0 \mid V,W,V^*=0)}\]
are estimated. The probability of $A$ can be estimated with data from $R=0$ using a logistic model. In the case of a randomized trial, the probability of $A$ is known by design and need not be estimated. However, estimating the probability of $A$ is expected to lead to a smaller asymptotic variance \citep{vdL2003}. Next, a model for $E[Y | A,V,W,R=0,V^*=0]$ is fit using weighted maximum likelihood using $\hat{\pi}(O)$ as the weights. Using this fitted model, predicted values of $Y$ with $a=1$, referred to as pseudo-outcomes and denoted by $\hat{Y}^1$, are generated for all observations with $R=1,V^*=0$. This process is repeated for $\hat{Y}^0$. To estimate the parameters of the MSM, the pseudo-outcomes for each $i$ with $R=1,V^*=0$ contribute twice, once for $\hat{Y}^1$ and once for $\hat{Y}^0$, to the score function of the specified MSM \citep{snowden_implementation_2011}. A point estimate for $\psi$ can then be computed by plugging in $\hat{\alpha}$ and $\nu$ into the synthesis MSM estimator (i.e., all observations with $R=1$ contribute). A proof of consistency and asymptotic normality for this estimator under correct specification of the mathematical model is provided in Appendix 2. Inference, where uncertainty in $\nu$ is incorporated, is considered in Section 3.1.3.

\subsubsection{Synthesis Estimator based on the Conditional Average Causal Effect}
Instead of modeling the mean under each value of $a$, an alternative approach models how the CACE varies by $V$. Specifically, consider the following re-expression of Equation (\ref{Eq6})
\[\psi = E[Y^1 - Y^0 \mid R=1] = E[E[Y^1 - Y^0 \mid V,R=1] \mid R=1].\]
This re-expression suggests an estimator based on modeling the CACE. The synthesis CACE estimator is
\[\widehat{\psi}_{CACE} = \frac{1}{n_1} \sum_{i=1}^{n} \left[ \mathcal{G}(O_i; \hat{\gamma}, \hat{\eta}, \delta) I(R_i = 1) \right]\]
where
\[\mathcal{G}(O_i; \hat{\gamma}, \hat{\eta}, \delta) = \mathpzc{s}(O_i; \hat{\gamma}, \hat{\eta}) + \mathpzc{m}(O_i; \delta) \]
with $\mathpzc{s}(\cdot)$ denoting a statistical CACE model with parameters $\gamma$ and nuisance parameters $\eta$, and $\mathpzc{m}(\cdot)$ denoting a mathematical model with $\lambda = \delta$ corresponding to shifts in the CACE. To illustrate, suppose the following CACE model specification,
\[E[Y^1 - Y^0 \mid V,R=1] = \gamma_0 + \gamma_1 V + V^*\{\delta_1 V + \delta_2 V^2\}\]
where the functional form for $V$ is allowed to vary between the statistical and mathematical models. Here, $\gamma_0$ is the causal effect among $V=0$ and $\gamma_1$ describes how the causal effect changes linearly with $V$ and $\delta$ modifies the relationship for values of $V$ above $v_0$. To summarize, this synthesis model relies on external information for how the CACE varies by $V$. \edit{This is in contrast to the MSM}, which requires specifying how \edit{$Y^a$ varies by $V$}.

Application of the synthesis CACE estimator proceeds in a similar way as the statistical MSM estimator up to the generation of the pseudo-outcomes. Instead the difference between the pseudo-outcomes, $\hat{Y}^1 - \hat{Y}^0$, is regressed on $V$ to estimate $\gamma$ using the data subset to $R=1,V^*=0$ \citep{van_der_laan_targeted_2015,kennedy_optimal_2020}. The point estimate for $\psi$ can then be computed by plugging in $\hat{\gamma}$ and $\delta$ in the CACE model. As before, inference is deferred until Section 3.1.3. Similarly, a proof of consistency and asymptotic normality for this estimator with correct specification of the mathematical model is provided in Appendix 2.

\subsubsection{Constructing a Mathematical Model}

External sources of information are used to build a mathematical model. Here, brief guidance is provided for obtaining appropriate external information and using it to build a mathematical model. General considerations for building mathematical models have been extensively discussed in the wider epidemiology and disease ecology literature \citep{krijkamp_microsimulation_2018,railsback_agent-based_2011,roberts_conceptualizing_2012,slayton_modeling_2020}, and the review here should only be taken as an overview.

The first step of the process is to compile what is known regarding the relationship between $A$ and $Y$. Commonly, this is done through literature reviews or studying previously developed mathematical models. When reviewing the literature, one should be open to considering diverse sources of external information (e.g., trials, observational studies, studies on exposure or treatments with similar mechanisms of action, pharmacokinetic studies, animal models). Additionally, a panel of subject-matter experts could be assembled to assess and summarize the current evidence base, following best practices for expert knowledge elicitation \citep{ohagan_expert_2019}. This approach has been used to elicit sensitivity analysis parameters \citep{shepherd_does_2008}. However obtained, the collected information should then be used to determine the structure of the mathematical model. For example, the review or panel may indicate that information on how $A,V,W$ interact with $Y^a$ is lacking, but interactions between intermediates are well understood (e.g., vaccine to antibody response to infection given exposure). In such cases, knowledge about the causal relationships between those intermediate mechanisms could be used to construct a mechanistic model \citep{kirkeby_practical_2021,lessler_mechanistic_2016}. Notice that the proposed estimators are all based on modifying the functional forms for nonpositive regions of a statistical model. While we believe modifying predictions from a statistical model for nonpositive regions to be the easiest to explain and interpret, the expression in (\ref{Eq6}) allows the mathematical model to be entirely separate from the statistical model. This means that in principle one can use any mathematical model in place of $\psi_1$. After determining the overall structure, possible values for the parameters of the mathematical model must be selected. When selecting values it must be remembered that use of a mathematical model is premised on Equation (\ref{Eq7}). Therefore, the choices of parameter values should accurately reflect the uncertainty of external information sources. Here, the Bayesian literature on prior elicitation could also be used to help guide mathematical model parameter selection \citep{mikkola_prior_2023,bockting_simulation-based_2023,icazatti_preliz_2023}. 

So far, inference has not been addressed. This was delayed until the description of building mathematical models since inference should proceed based on the mathematical model. Here, two perspectives on inference are provided, with the differences arising from how parameters of the mathematical models are specified.

First, in building the mathematical model one may find that the only reliable external information is regarding the bounds or extremes of the possible values for $\lambda$. In this case, one may search across the combinations of $\lambda$ that provide the smallest and largest point estimates for $\psi$. Then, one can report the range of $\psi$, without assigning differing plausibilities between values in the range. This output can be be viewed as bounds on $\psi$ \citep{frechet_generalisation_1935,manski_nonparametric_1990,cole_nonparametric_2019}, narrowed by integration of external information. Wald confidence intervals for each of the boundaries can then be computed using the empirical sandwich variance estimator with estimating equations for the corresponding synthesis estimator and plugging in the $\lambda$ combinations resulting in the extremes \citep{stefanski_calculus_2002,zivich_delicatessen_2022}. Following the conceptual framework of \citep{vansteelandt_ignorance_2006}, the region within the bounds can be viewed as `ignorance' error (i.e., the inability or unwillingness to further delineate between values in the chosen range for $\lambda$), while the confidence intervals indicate the sampling error of the ignorance region.

Alternatively, one may find that the external information is rich enough to be able to specify distributions for $\lambda$ based on their corresponding plausibility. Here, a semiparametric bootstrap procedure is used \citep{zivich_transportability_2023}. First, the statistical parameters and the covariance matrix are estimated using the observed data. Next, parameters for the statistical model (i.e., $\alpha$ or $\gamma$) are drawn from a multivariate normal and parameters for the mathematical model are randomly drawn from their specified distributions. Using a random sample with replacement from $R=1$, $\hat{\psi}$ is computed using the random draw of the statistical and mathematical parameters. The resulting estimate is then saved. The process of randomly drawing the model parameters and sampling with replacement from $R=1$ is repeated, perhaps thousands of times. The collection of estimates can then be summarized numerically (e.g., percentiles) or visually (e.g., histograms). These results can then be interpreted under a semi-Bayesian framework, as done with quantitative bias analyses and related settings \citep{good_bayesnon-bayes_1992,greenland_bayesian_2006,fox_introduction_2021}. Alternatively, one could opt for alternative computational methods, like Markov Chain Monte Carlo, for estimation. Finally, if one also puts priors on the statistical model nuisance parameters, the resulting estimates can be interpreted under a Bayesian paradigm. 

\subsection{Other Approaches}

Having reviewed the proposed approach, attention is now turned to more standard approaches to dealing with structural sampling positivity violations. Three options are reviewed: restricting the target population, restricting the covariate set for sampling exchangeability, and extrapolating from a statistical model. 

\subsubsection{Restrict the Target Population}

As in the binary case \citep{zivich_transportability_2023}, the target population can be restricted to the regions where positivity holds. This solves nonpositivity by restricting inference to $\psi_0$. As shown, $\psi_0$ is identified following Equations (\ref{Eq1})-(\ref{Eq3}), (\ref{Eq4a}), and (\ref{Eq5a}). Estimation under these identification assumptions can then be accomplished with existing g-computation, IPW, or AIPW estimators for transportability \citep{westreich_transportability_2017,dahabreh_extending_2020}.

While valid, this approach addresses a different scientific question, as it provides inference for only a fraction of the original target population. If one were to presume, intentionally or mistakenly, that $\psi = \psi_0$, then one is implicitly assuming homogeneity of the causal effect between the positive and nonpositive regions. Even when $\psi_0$ is correctly interpreted, often decisions must be still made for those with $V > v_0$ (i.e., abstaining from a decision is a decision). Depending on how $A$ is distributed in the target population, harms can be inflicted. For example, if results led to policies or decisions that withheld $A$ from those with $V > v_0$ but $A$ would have been beneficial, then an analytical decision would have led to worse outcomes. Alternatively, if effects were to reverse for $V > v_0$, then offering $A$ would be actively harmful. By restricting to regions with positivity, one claims complete ignorance between these cases and offers no guidance for decision makers.

\subsubsection{Restrict the Covariate Set for Exchangeability}

Another option is to reduce the adjustment set considered. Rather than account for both $V,W$, the analysis could instead account for only $W$. This addresses Equation (\ref{Eq5}) by asserting effects are homogeneous over $V$ on the difference scale \citep{webster-clark_directed_2021}, so it can safely be ignored. Equations (\ref{Eq4}) and (\ref{Eq5}) then become
\begin{equation}
	\begin{split}
		E[Y^a \mid W=w,R=1] = & E[Y^a \mid W=w,R=0] \\ 
		&\text{ where } f_{W,R}(w,R=1) > 0
	\end{split}
	\tag{4"}
	\label{Eq4b}
\end{equation}
\begin{equation}
	\Pr(R=0 \mid W=w) > \epsilon > 0 \text{ where } f_{W,R}(w,R=1) > 0
	\tag{5"}
	\label{Eq5b}
\end{equation}
Similar to the restricted target population approach, g-computation, IPW, or AIPW estimators can be readily applied. While restricting the covariate set addresses the motivating question, it does so via a seemingly arbitrary, data-driven decision. The plausibility of these revised assumptions will often be suspect in practical applications, \edit{and the corresponding estimator can be substantially biased when there is additive effect heterogeneity by $V$.}

Notice the relationship between this approach and the others considered thus far. In the binary setting, restricting the target population and assuming $\psi = \psi_0$ can be shown to be a special case of restricting the covariate set \citep{zivich_transportability_2023}. However, this is not generally true in the continuous setting. The difference arises out of the previous approach being able to account for $V$ in the restricted population, whereas in the binary setting neither approach can incorporate the binary covariate due to nonpositivity. Similarly, restricting the covariate set was previously shown to be a special case of the synthesis approach in the context of nonpositivity with a binary covariate \citep{zivich_transportability_2023}. Again, this relationship is not generally true for nonpositivity with a continuous covariate for similar reasons.

\subsubsection{Statistical Model Extrapolation}

Rather than restrict either the target population or the adjustment set, one can use a statistical model fit to the $R=0$ data to \textit{extrapolate} to the full target population. Briefly, one would estimate the parameters of an outcome model conditional on $A,V,W$ using data from $R=0$ and then generate pseudo-outcomes for the full $R=1$ population. To enable extrapolation to the full target population, this approach abandons \textit{nonparametric} identification and instead opts for \textit{parametric} (or local) identification \citep{king2006dangers}. To extrapolate to nonpositive regions, some statistical model for the outcome process must be imposed. The simplest case is g-computation, where a model for $Y$ given $A,V,W$ in the \edit{external} population is fit and then used to compute $\psi$. IPW is less straightforward, as there are no individuals in the \edit{external} population capable of standing-in for the units with $V^*=1$. Another way to view this issue is to note that the IPW estimator fits a nonparametric model for the outcome process. As a result, the IPW estimator cannot be used directly to extrapolate. However, one could use IPW to estimate a MSM using $Y$ given $A,V$ with $R=0$ and then use that estimated MSM to extrapolate to the target population. Notice that this replaces the nonparametric outcome model with a parametric model for $V$ to allow for the extrapolations. When using weighted regression AIPW \citep{vansteelandt_invited_2011}, the outcome model can be used to extrapolate to the full target population. 

A related, but distinct, type of positivity violations are \textit{random} positivity violations, where the probability conditional on $V,W$ is non-zero for the population but some $v,w$ combination was not observed in a finite sample of $n$ units \citep{zivich_positivity_2022,westreich_invited_2010}. While parametric models are often used to address random positivity violations \citep{zhu_core_2021}, a problem occurs in their use for \textit{structural} positivity violations considered thus far. To allows for the use of a parametric model for estimation, an assumption of correct model specification is made in addition to Equations (\ref{Eq1})-(\ref{Eq3}). Correct model specification, for a set of probability distributions governed by the parameters $\eta$ denoted as $\mathcal{M}_\eta$, can be expressed as 
\[\Pr(A=a \mid V,W,R=0) \in \mathcal{M}_{\eta_1}\]
\[\Pr(R=1 \mid V,W) \in \mathcal{M}_{\eta_2}\]
for the propensity score and sampling models, respectively, or 
\[E[Y \mid A,V,W,R=0] \in \mathcal{M}_{\eta_3}\]
for the outcome model. In words, this assumption states that the set of probability distributions for each model contains the corresponding true distribution. To enable extrapolation from the \edit{external} population to the target population for structural nonpositivity, one must make the corresponding model specification assumption(s) and then further assume that the outcome model used to extrapolate is correctly specified \textit{for the nonpositive regions} \citep{king2006dangers}. To see the distinction here, note that one could correctly specify the model in $R=0$ (i.e., the model is correctly specified for all $V \le v_0$). However, the extrapolations from that model could still be incorrect for $V > v_0$ (Figure \ref{Fig1}). 

To clarify this distinction further, notice that statistical model extrapolation is a special case of synthesis modeling. For didactic purposes, suppose a statistical MSM is used for extrapolation and is specified to be 
\[E[Y^a | V,R=1] = \alpha_0 + \alpha_1 a + \alpha_2 V + \alpha_3 a V\]
which is equivalent to 
\[E[Y^a | V,R=1] = \alpha_0 + \alpha_1 a + \alpha_2 V + \alpha_3 a V + m(a, O_i; \nu)\]
when $m(a, O; \nu) = 0$ for all $a \in \{0,1\}$ and $f(v,w \mid R=0) > 0$. In words, the mathematical model always contributes exactly zero. So, use of extrapolation is equivalent to assuming that the true mathematical model is the same as the estimated statistical model, with certainty. When viewed in this manner, the extremeness of the assumption for extrapolation from a statistical model may become clearer. It also suggests that even if one were to avoid using synthesis modeling because they believe there to be limited or weak external knowledge, then one must also recognize that extrapolation is similarly unjustified. Finally, it suggests that users of the extrapolation approach can incorporate uncertainty by using a synthesis model with distributions centered at zero. 

\section{Simulations}

To explore the performance of the proposed estimators, a simulation study was conducted. The simulation design was inspired by trials on ART for persons with HIV that had inclusion criteria based on pre-randomization CD4 cell counts \citep{hammer_trial_1996,hammer_controlled_1997}. 

\subsection{Data Generating Mechanisms}

Let $A_i$ be assigned ART (1: yes, 0: no) and $Y_i$ be the outcome of CD4 cell count at one-year following randomization. Further, let $V_i$ be CD4 cell count before randomization and $W_i$ indicate gender (1: female, 0: male). The parameter of interest was the average causal effect in the target population, $\psi$. 

The distribution of baseline variables in the target population (i.e., $R=1$) was $V \sim 375 \times \text{Weibull}(\vartheta = 1, \omega=1.5)$, with the probability density function parameterized as $\frac{\omega}{\vartheta} \left(\frac{t}{\vartheta}\right)^{\omega-1} \exp\left( -\left\{ \frac{t}{\vartheta} \right\}^\omega \right) $, and $W \sim \text{Bernoulli}(0.2)$. The \edit{external} population (i.e., $R=0$) observations were selected from the target population using the following sampling model
\begin{equation*}
	\Pr(R=0 \mid V,W) = 
	\begin{cases}
		\text{expit}(-0.02V + 2W) & V \le 300 \\
		0 						  & V > 300		
	\end{cases}
\end{equation*}
Therefore, $V^* = I(V > 300)$ and structural positivity for transporting between populations was violated. Within the \edit{external} population, $\Pr(A=1 | R=0) = 0.5$. Two scenarios were considered. In the first scenario, the outcome model was such that linear extrapolations to the nonpositive region were valid,
\[Y_i^a = -20 + 70a + V_i + 0.12a V_i - 2W_i + 5aW_i + \epsilon_i\]
where $\epsilon \sim \text{Normal}(\mu=0,\sigma=25)$. In the second scenario, the relationship between $Y^a$ and $V$ was varied in the nonpositive region, such that the linear extrapolations from the statistical model were invalid. Specifically, the one-year increase in CD4 was less among those with a higher CD4 at baseline,
\begin{equation*}
	\begin{split}
		Y_i^a = & -20 + 70a + V_i + 0.2a V_i - 2W_i + 5aW_i \\
		& - 0.2a \{V_i - 300\} I(V_i > 300) - 0.3a \{V_i - 800\} I(V_i > 800) + \epsilon_i.
	\end{split}
\end{equation*}
To reflect how CD4 cell counts are measured in practice, values were rounded to integers for both mechanisms. To prevent logically inconsistent for CD4, values of $Y_i^a$ below one were set to one. Sample sizes of $n_1 = n_0 = 1000$ and $n_1 = 1000, n_0 = 500$ were considered, with each scenario consisting of 1000 repetitions. 

To evaluate the different approaches, the following metrics were considered: bias, relative bias, CI width, and 95\% CI coverage \citep{morris_using_2019}. Bias was defined as the mean of the difference between $\hat{\psi}$ for each method versus $\psi$. \edit{Here, $\psi$ was} approximated \edit{for each scenario by separately} simulating potential outcomes for 20 million observations from \edit{the corresponding outcome model for} the target population. Relative bias was defined as $\frac{\hat{\psi} - \psi}{\psi}$. CI width was reported as the mean difference between the upper and lower values of the CIs, and used as a measure of precision. Finally, 95\% CI coverage was defined as the proportion of 95\% CIs that covered $\psi$. Simulations were conducted using Python 3.9.6 (Beaverton, OR, USA) with the following packages: \texttt{NumPy} \citep{harris_array_2020}, \texttt{SciPy} \citep{virtanen_scipy_2020}, \texttt{pandas} \citep{mckinney_data_2010}, and \texttt{delicatessen} \citep{zivich_delicatessen_2022}. Code to replicate the simulations is available at https://github.com/pzivich/publications-code.

\subsection{Non-synthesis Estimators}

Estimators for the restricted target population, restricted covariate set, and extrapolation approaches were all implemented using a weighted regression AIPW estimator. The weighted regression AIPW estimator was expressed as an M-estimator \citep{stefanski_calculus_2002}, where an M-estimator, $\hat{\theta}$, is the solution to the estimating equation, $\sum_{i=1}^{n} \phi(O_i, \theta) = 0$ with the estimating function $\phi(O_i, \theta)$. The variance was estimated using the empirical sandwich variance estimator \citep{stefanski_calculus_2002,zivich_delicatessen_2022}. The stacked estimating functions for the weighted regression AIPW estimators were
\begin{equation*}
	\phi(O_i; \theta) = 
	\begin{bmatrix}
		\phi_{\eta_1} \\ 
		\phi_{\eta_2} \\ 
		\phi_{\eta_3} \\ 
		\phi_{\mu_c^1} \\ 
		\phi_{\mu_c^0} \\ 
		\phi_{\psi_c} \\ 		
	\end{bmatrix}
	=
	\begin{bmatrix}
		(1 - R_i) \left\{ A_i - \text{expit}\left(\mathbb{Z}_i \eta_1^T\right) \right\} \mathbb{Z}_i^T \\
		\left\{ R_i - \text{expit}\left(\mathbb{U}_i \eta_2^T\right) \right\}\mathbb{U}_i^T \\	
		(1 - R_i) \pi(A_i, V_i, W_i; \eta_1, \eta_2) \left\{ Y_i - \mathbb{X}_i \eta_3^T \right\}\mathbb{X}_i^T \\	
		R_i \left[ \hat{Y}^1_i - \mu_c^1 \right] \\ 
		R_i \left[ \hat{Y}^0_i - \mu_c^0 \right] \\ 
		(\mu_c^1 - \mu_c^0) - \psi_c
	\end{bmatrix}
\end{equation*}
where $\theta = (\eta_1, \eta_2, \eta_3, \mu_c^1, \mu_c^0, \psi_c)$,
\[ \pi(A_i, V_i, W_i; \eta_1, \eta_2) = \frac{1 - \text{expit}\left(\mathbb{U}_i \eta_2^T\right)}{\text{expit}\left(\mathbb{U}_i \eta_1^T\right)} \times \frac{I(R_i = 0)}{A_i 
\text{expit}\left(\mathbb{Z}_i \eta_1^T\right)
+ (1 - A_i) \left\{1 - \text{expit}\left(\mathbb{Z}_i \eta_1^T\right) \right\} }, \]
$\mathbb{Z,U,X}$ are the design matrices for the observed data, and $\hat{Y}^a = \mathbb{X}_i(a) \hat{\eta}^T$ where $\mathbb{X}(a)$ is the same design matrix as $\mathbb{X}$ but with $A$ set to $a$. The estimating function $\phi_{\eta_1}$ is a logistic regression model for the probability of $A$ in the \edit{external} population conditional on variables in $\mathbb{Z}$, $\phi_{\eta_2}$ is a logistic regression model for the probability of $R$ conditional on variables in $\mathbb{U}$, and $\phi_{\eta_3}$ is an inverse probability weighted linear regression model for the outcome in the \edit{external} population conditional on $\mathbb{X}$. $\phi_{\mu_c^1}$ and $\phi_{\mu_c^0}$ are the mean under $A=1$ and $A=0$ in the target population, respectively. Finally, $\phi_{\psi_c}$ is for the parameter of interest. 

The stacked estimating functions are the same for the three non-synthesis approaches, with the distinction coming from what observations are used in each estimating equation and the specification of the design matrices. For the restricted target population, the data were restricted to observations where $V_i^* = 0$ and the design matrices were $\mathbb{Z}= 1$, $\mathbb{U} = (1, V, W)$, and $\mathbb{X} = (1, A, V, W, AV, AW)$. For the restricted covariate set, the full data was used but the design matrices were instead $\mathbb{Z}= 1$, $\mathbb{U} = (1, W)$, and $\mathbb{X} = (1, A, W, AW)$. Finally, the extrapolation approach used the full population with the same design matrices as the restricted population approach. However, the sampling model was fit using only $V^* = 0$ (i.e., $(1-V^*) \times \phi_{\eta_2}$).

\subsection{Synthesis Estimators}

The described synthesis estimators were implemented by expressing them as estimating equations. The corresponding estimating functions for the synthesis MSM were
\begin{equation}
	\phi(O_i; \theta) 
	= 
	\begin{bmatrix}
		\phi_{\eta_1} \\ 
		\phi_{\eta_2} \\ 
		\phi_{\eta_3} \\ 
		\phi_{\alpha} \\ 		
		\phi_{\mu_s^1} \\ 
		\phi_{\mu_s^0} \\ 
		\phi_{\psi_{sm}} \\ 	
	\end{bmatrix}
	=
	\begin{bmatrix}
		(1 - R_i) \left\{ A_i - \text{expit}\left(\mathbb{Z}_i \eta_1^T\right) \right\} \mathbb{Z}_i^T \\
		(1 - V_i^*) \left\{ R_i - \text{expit}\left(\mathbb{U}_i \eta_2^T\right) \right\}\mathbb{U}_i^T \\	
		(1- R_i) \pi(A_i, V_i, W_i; \eta_1, \eta_2) \left\{ Y_i - \mathbb{X}_i \eta_3^T \right\}\mathbb{X}_i^T \\
		(1 - V_i^*) R_i \left[ \left( \hat{Y}_i^1 - \mathbb{W}_i(1) \alpha^T \right) \mathbb{W}_i(1)^T +  \left( \hat{Y}_i^0 - \mathbb{W}_i(0) \alpha^T \right) \mathbb{W}_i(0)^T \right] \\	
		R_i \left[ \mathbb{W}_i(1) \alpha^T + \mathbb{W}^*_i(1) \nu^T - \mu_s^1 \right] \\ 
		R_i \left[ \mathbb{W}_i(0) \alpha^T + \mathbb{W}^*_i(0) \nu^T - \mu_s^0 \right] \\ 
		(\mu_s^1 - \mu_s^0) - \psi_{sm}
	\end{bmatrix}
	\label{SynMSM}
\end{equation}
where $\mathbb{W}(a)$ is the design matrix for the positive region of the MSM with $A$ set as $a$, and $\mathbb{W}^*(a)$ is the design matrix for the nonpositive region of the MSM with $A$ set as $a$. Again, $\phi_{\eta_1}, \phi_{\eta_2}, \phi_{\eta_3}$ are the nuisance models for the statistical AIPW estimator. $\phi_{\alpha}$ is for the MSM parameters, which adds the contributions of participant $i$ under each value of $a$. This estimating function is equivalent to the stacking approach described in Snowden et al. \citep{snowden_implementation_2011}. $\phi_{\mu_s^1},\phi_{\mu_s^0}$ are the mean under $A=1$ and $A=0$ of the target population using predictions from the synthesis model. $\phi_{\psi_{sm}}$ is for the parameter of interest.

The specified design matrices were $\mathbb{Z}= 1$, $\mathbb{U} = (1, V, W)$, $\mathbb{X} = (1, A, V, W, AV, AW)$, $\mathbb{W} = (1, A, V, AV)$, and $\mathbb{W}^* = (A\{V-300\} I(V>300), A \{V-800\} I(V>800))$. This corresponds to a synthesis MSM of 
\begin{equation*}
	\begin{split}
		\mathcal{F}_a(O_i; \alpha, \eta, \beta) = & \alpha_0 + \alpha_1 a + \alpha_2 V_i + \alpha_3 a V_i \\
		& + \nu_1 a \{V_i - 300\} I (V_i > 300) + \nu_2 a \{V_i - 800\} I (V_i > 800).
	\end{split}	
\end{equation*}
Here, $\nu$ is a constant in the estimating function (i.e., it is not estimated). As the semiparametric bootstrap was used for inference in the simulations, only the estimating functions up to $\phi_{\alpha}$ were needed. The $\widehat{Cov}(\hat{\alpha})$ was estimated via the empirical sandwich variance estimator. Uncertainty in the sample of the target was incorporated by resampling with replacement and uncertainty in the parameters was incorporated via random draws from the corresponding distributions. The semiparametric bootstrap was repeated 10,000 times. The point estimate was the median of the collection of point estimates, and 95\% confidence intervals (CIs) were obtained using the 2.5\textsuperscript{th} and 97.5\textsuperscript{th} percentiles.

The estimating functions for the synthesis CACE estimator were
\begin{equation}
	\phi(O_i; \theta) 
	= 
	\begin{bmatrix}
		\phi_{\eta_1} \\ 
		\phi_{\eta_2} \\ 
		\phi_{\eta_3} \\ 
		\phi_{\gamma} \\ 		
		\phi_{\psi_{sc}} \\ 	
	\end{bmatrix}
	=
	\begin{bmatrix}
		(1 - R_i) \left\{ A_i - \text{expit}\left(\mathbb{Z}_i \eta_1^T\right) \right\} \mathbb{Z}_i^T \\
		(1 - V_i^*) \left\{ R_i - \text{expit}\left(\mathbb{U}_i \eta_2^T\right) \right\}\mathbb{U}_i^T \\	
		(1- R_i) \pi(A_i, V_i, W_i; \eta_1, \eta_2) \left\{ Y_i - \mathbb{X}_i \eta_3^T \right\}\mathbb{X}_i^T \\	
		(1 - V_i^*) R_i \left\{ \left(\hat{Y}_i^1 - \hat{Y}_i^0\right) - \mathbb{V}_i \gamma^T \right\} \\
		R_i \left[ \mathbb{V}_i \gamma^T + \mathbb{V}_i^* \delta^T - \psi_{sc} \right] \\ 
	\end{bmatrix}
	\label{SynCACE}
\end{equation}
where $\mathbb{V}$ is the design matrix for the positive region of the CACE model, and $\mathbb{V}^*$ is the design matrix for the nonpositive region of the CACE model. The design matrices for the statistical models were $\mathbb{Z}= 1$, $\mathbb{U} = (1, V, W)$, $\mathbb{X} = (1, A, V, W, AV, AW)$, and $\mathbb{V} = (1, V)$. The design matrix for the mathematical model contributions was $\mathbb{V}^* = (\{V-300\} I(V>300), \{V-800\} I(V>800))$. Therefore, the specified CACE model was
\[\mathcal{G}(O_i; \gamma, \eta, \delta) = \gamma_0 + \gamma_1 V + \delta_1 \{V_i - 300\} I(V_i > 300) + \delta_2 \{V_i - 800\} I(V_i > 800).\]
As with the synthesis MSM estimator, uncertainty was incorporated using the semiparametric bootstrap with 10,000 iterations. Similarly, only the estimating functions up to $\phi_{\gamma}$ were needed to implement the semiparametric bootstrap. 

Since data were simulated, the following process was used to generate external information for the mathematical models. Separate from the previous observations, $n_2$ observations were created from the target population distribution. For these observations, $A_i$ was randomly assigned and outcomes were generated from the corresponding model for the scenario. These observations were used to estimate the MSM
\[E[Y^a | V] = \alpha_0 + \alpha_1 a + \alpha_2 V + \alpha_3 a V + \nu_1 a \{V - 300\} I(V>300) + \nu_2 a \{V - 800\} I(V>800)\]
and CACE model,
\[E[Y^1 - Y^0 | V] = \gamma_0 + \gamma_1 V + \delta_1 \{V - 300\} I(V>300) + \delta_2 \{V - 800\} I(V>800).\]
The estimated $\lambda$ values from these models were then used as the external information for the corresponding mathematical models. This approach means that the input mathematical parameters are correctly specified in expectation but vary between iterations. Further, the precision of the external information can be modified by increasing $n_2$. 

Several variations in the specification of the mathematical model parameters were considered. First, independent normal distributions with the corresponding point and variance estimates were considered for each parameter with $n_2 = 2000$. To examine how different distributions modified the results with $n_2=2000$, independent trapezoidal distributions ranging from $\hat{\lambda} \pm 3 \sqrt{\widehat{Var}(\hat{\lambda})}$ with a shelf from $\hat{\lambda} \pm \sqrt{\widehat{Var}(\hat{\lambda})}$ were considered. To study how precision of the external information modified results, independent normal distributions were again considered but with $n_2=8000$. Finally, an uncertain null distribution was implemented with a uniform distribution from $-0.3$ to $0.3$ to compare to the extrapolation approach.

\subsection{Results}

Performance of the estimators varied as expected. In scenario one, the restricted target population and restricted covariate set were substantially biased and had 0\% CI coverage for the full target population with $n_0 = 1000$ (Table \ref{Tab1}). As expected in this scenario, extrapolation had little relative bias and CI coverage was near expected levels. All the synthesis estimators similarly had little relative bias, with CI coverage near 95\% for all mathematical model specifications besides the uniform null. The uniform null had 100\% coverage. Regarding precision, the synthesis estimators with uniform null mathematical model specification had the widest CI. Extrapolation and synthesis estimators with other mathematical model specifications had similar precision, with slightly better precision for the synthesis estimators. Comparing the synthesis estimators to the independent normal distributions based on $n_2 = 2000$, the trapezoidal specification used here was slightly less precise. When the precision of the specified mathematical model parameters was increased for the independent normal (i.e., $n_2 = 8000$), the precision for the parameter of interest was only slightly improved. This gain in precision was less than might be expected, as the sample size used to generate this external information was four times as large as the other independent normal specification. However, it must be remembered that the choice of the mathematical model parameters only affect observations in the nonpositive regions (i.e., it does not affect the uncertainty in the statistical model). For $n_0 = 500$, the same general patterns held. However, coverage was below nominal for the extrapolation approach (Table \ref{Tab1}). Further, the precision of the synthesis approaches was generally improved over the extrapolation approach. Both results were likely attributable to the occurrence of finite-sample bias resulting in more extreme extrapolations from statistical model.

\begin{table}[h]
	\caption{Simulation results for a linear extension of the relationship between baseline and subsequent CD4 cell counts}
	\centering
	\begin{tabular}{lllcccc} 
		\hline
		&  &                                             & Bias  & Relative bias & CLD  & Coverage  \\ 
		\cline{4-7}
		\multicolumn{3}{l}{$n_1=1000,n_0=1000$}       &       &               &      &           \\
		& \multicolumn{2}{l}{Restricted Population}      & -21.0 & -0.19         & 7.7  & 0\%       \\
		& \multicolumn{2}{l}{Restricted Covariates}      & -26.5 & -0.24         & 18.5 & 0\%       \\
		& \multicolumn{2}{l}{Extrapolation}              & 1.5   & 0.01          & 27.5 & 95\%      \\
		& \multicolumn{2}{l}{Synthesis MSM}              &       &               &      &           \\
		&  & Normal ($n_2 = 2000$)                    & 2.4   & 0.02          & 25.0 & 94\%      \\
		&  & Trapezoid ($n_2 = 2000$)                 & 2.4   & 0.02          & 25.6 & 95\%      \\
		&  & Normal ($n_2 = 8000$)                    & 2.3   & 0.02          & 24.3 & 93\%      \\
		&  & Uniform Null                                & 2.3   & 0.02          & 73.2 & 100\%     \\
		& \multicolumn{2}{l}{Synthesis CACE}             &       &               &      &           \\
		&  & Normal ($n_2 = 2000$)                    & 2.5   & 0.02          & 25.0 & 93\%      \\
		&  & Trapezoid ($n_2 = 2000$)                 & 2.5   & 0.02          & 25.7 & 94\%      \\
		&  & Normal ($n_2 = 8000$)                    & 2.3   & 0.02          & 24.3 & 94\%      \\
		&  & Uniform Null                                & 2.3   & 0.02          & 73.3 & 100\%     \\
		\multicolumn{3}{l}{$n_1 = 1000, n_0 = 500$}   &       &               &      &           \\
		& \multicolumn{2}{l}{Restricted Population}      & -21.0 & -0.19         & 10.6 & 0\%       \\
		& \multicolumn{2}{l}{Restricted Covariates}      & -26.7 & -0.24         & 26.1 & 2\%       \\
		& \multicolumn{2}{l}{Extrapolation}              & 2.2   & 0.02          & 36.7 & 93\%      \\
		& \multicolumn{2}{l}{Synthesis MSM}              &       &               &      &           \\
		&  & Normal ($n_2 = 2000$)                    & 2.5   & 0.02          & 33.5 & 94\%      \\
		&  & Trapezoid ($n_2 = 2000$)                 & 2.4   & 0.02          & 34.0 & 94\%      \\
		&  & Normal ($n_2 = 8000$)                    & 2.4   & 0.02          & 33.0 & 93\%      \\
		&  & Uniform Null                                & 2.4   & 0.02          & 77.3 & 100\%     \\
		& \multicolumn{2}{l}{Synthesis CACE}             &       &               &      &           \\
		&  & Normal ($n_2 = 2000$)                    & 2.5   & 0.02          & 33.6 & 93\%      \\
		&  & Trapezoid ($n_2 = 2000$)                 & 2.5   & 0.02          & 34.0 & 94\%      \\
		&  & Normal ($n_2 = 8000$)                    & 2.4   & 0.02          & 33.0 & 93\%      \\
		&  & Uniform Null                             & 2.4   & 0.02          & 77.3 & 100\%     \\
		\hline
	\end{tabular}
	\floatfoot{MSM: Marginal Structural Model, CACE: Conditional Average Causal Effect, CLD: Confidence Limit Difference.
	}
	\label{Tab1}
\end{table}

In scenario two, each of the non-synthesis approaches had substantial bias with $n_0 = 1000$ (Table \ref{Tab2}). Both the restricted target population and restricted covariate set were downwardly biased for estimation of $\psi$. Extrapolation instead overestimated the effect of ART on CD4. CI coverage was near zero for all the non-synthesis approaches. The synthesis estimators fared better. Both the synthesis MSM and CACE estimators that used uniform null distributions had similar amounts of bias to the extrapolation approach. However, the uncertainty expressed in those parameters decreased the precision of the results and also resulted in greater CI coverage. For the mathematical model parameters that were correctly specified in expectation, relative bias of the synthesis estimators was near zero. For the independent normal distribution based on $n_2 = 2000$, CI coverage was approximately 95\%. When using a trapezoidal distribution for the mathematical model parameters, coverage was similar but precision decreased. When mathematical model parameters were set to be more precise, precision for estimating the parameter of interest was only slightly improved. Again, this is related to the decomposition in (\ref{Eq6}) and the mathematical model only influencing precision in $\psi_1$. For $n_0 = 500$, precision was decreased for all estimators but the pattern of results remained similar otherwise. 

\begin{table}
	\caption{Simulation results for a non-linear extension of the relationship between baseline and subsequent CD4 cell counts }
	\centering
	\begin{tabular}{lllcccc} 
		\hline
		&  &                                             & Bias  & Relative bias & CLD  & Coverage  \\ 
		\cline{4-7}
		\multicolumn{3}{l}{$n_1=1000,n_0=1000$}       &       &               &      &           \\
		& \multicolumn{2}{l}{Restricted Population}      & -11.2 & -0.10         & 8.0  & 0\%       \\
		& \multicolumn{2}{l}{Restricted Covariates}      & -20.0 & -0.17         & 19.1 & 2\%       \\
		& \multicolumn{2}{l}{Extrapolation}              & 25.3  & 0.22          & 28.1 & 7\%       \\
		& \multicolumn{2}{l}{Synthesis MSM}              &       &               &      &           \\
		&  & Normal ($n_2 = 2000$)                    & 1.0   & 0.01          & 25.0 & 95\%      \\
		&  & Trapezoid ($n_2 = 2000$)                 & 1.0   & 0.01          & 25.6 & 95\%      \\
		&  & Normal ($n_2 = 8000$)                    & 1.0   & 0.01          & 24.2 & 94\%      \\
		&  & Uniform Null                                & 26.1  & 0.23          & 73.2 & 95\%      \\
		& \multicolumn{2}{l}{Synthesis CACE}             &       &               &      &           \\
		&  & Normal ($n_2 = 2000$)                    & 1.1   & 0.01          & 25.0 & 94\%      \\
		&  & Trapezoid ($n_2 = 2000$)                 & 1.1   & 0.01          & 25.6 & 95\%      \\
		&  & Normal ($n_2 = 8000$)                    & 1.0   & 0.01          & 24.2 & 94\%      \\
		&  & Uniform Null                                & 26.1  & 0.23          & 73.2 & 95\%      \\
		\multicolumn{3}{l}{$n_1 = 1000, n_0 = 500$}   &       &               &      &           \\
		& \multicolumn{2}{l}{Restricted Population}      & -11.3 & -0.10         & 10.8 & 2\%       \\
		& \multicolumn{2}{l}{Restricted Covariates}      & -20.2 & -0.18         & 27.0 & 17\%      \\
		& \multicolumn{2}{l}{Extrapolation}              & 25.4  & 0.22          & 37.0 & 25\%      \\
		& \multicolumn{2}{l}{Synthesis MSM}              &       &               &      &           \\
		&  & Normal ($n_2 = 2000$)                    & 0.9   & 0.01          & 33.5 & 94\%      \\
		&  & Trapezoid ($n_2 = 2000$)                 & 0.8   & 0.01          & 34.0 & 94\%      \\
		&  & Normal ($n_2 = 8000$)                    & 0.9   & 0.01          & 32.9 & 93\%      \\
		&  & Uniform Null                                & 25.9  & 0.23          & 77.5 & 92\%      \\
		& \multicolumn{2}{l}{Synthesis CACE}             &       &               &      &           \\
		&  & Normal ($n_2 = 2000$)                    & 0.9   & 0.01          & 33.5 & 94\%      \\
		&  & Trapezoid ($n_2 = 2000$)                 & 0.9   & 0.01          & 34.0 & 94\%      \\
		&  & Normal ($n_2 = 8000$)                    & 0.9   & 0.01          & 32.9 & 93\%      \\
		&  & Uniform Null                                & 25.9  & 0.23          & 77.5 & 92\%      \\
		\hline
	\end{tabular}
	\floatfoot{MSM: Marginal Structural Model, CACE: Conditional Average Causal Effect, CLD: Confidence Limit Difference.
	}
	\label{Tab2}
\end{table}

\section{Application}

For illustration, we adapted a \edit{previously published transportability} example in HIV research \citep{dahabreh_sensitivity_2023}. We aimed to quantify the short-term effectiveness of two-drug versus one-drug ART on subsequent CD4 T cell counts per mm\textsuperscript{3} among HIV infected women in the United States (US). Data was available from the AIDS Clinical Trial Group (ACTG) 175 randomized trial and wave one of the Women's Interagency HIV study (WIHS) cohort. The parameter of interest is the mean difference in CD4 cell count at 20 weeks given assignment to two-drug ART versus assignment to one-drug ART in the WIHS target population.

\subsection{Data Sources}

ACTG 175 was a double-masked randomized trial comparing different ART regimens to prevent AIDS, death, and decline in CD4 among HIV-1 infected participants \citep{hammer_trial_1996}. Individuals were recruited from the US between December 1991 and October 1992. Eligibility criteria included 12 years of age or older, no history of previous AIDS-defining illness, a Karnofsky score of at least 70, and screening CD4 counts between 200 and 500 cells/mm\textsuperscript{3}. In the trial, participants were randomized to one of four ART regimes.  For the presented analysis, trial arms are restricted to zidovudine-didanosine or zidovudine-zalcitabine (combined as two-drug ART), and zidovudine (one-drug ART). Covariates included baseline CD4 (cells/mm\textsuperscript{3}), age (years), race (white, non-white), gender (male, female), and weight (kilograms). The outcome was CD4 measured at 20$\pm$5 weeks. \edit{Note that the baseline CD4 count was not necessarily equal to the screening CD4 count, as baseline CD4 was an average of two measurements prior to treatment excluding the screening CD4. The screening CD4 values were not available in the public-use ACTG 175 data set, so the available CD4 counts were not necessarily between 200 and 500.}

WIHS is an on-going US-based cohort study enrolling women with HIV and seronegative women \citep{barkan_womens_1998}. Here, wave one of participants with HIV, enrolled between October 1994 and November 1995, were considered. As WIHS did not restrict eligibility by baseline CD4, the women enrolled in WIHS are thought to be better representative of the desired target population (i.e., HIV infected women living in the US). Baseline variables included CD4, age, race (re-categorized as white, non-white), and weight. No information on ART or outcomes from WIHS were considered in the analysis. 

\begin{figure}
	\centering
	\caption {Baseline CD4 cell counts by data source (A) and the estimated conditional average causal effect for two-drug versus one-drug antiretroviral therapy in the WIHS target population (B)}
	\includegraphics[width=0.85\linewidth]{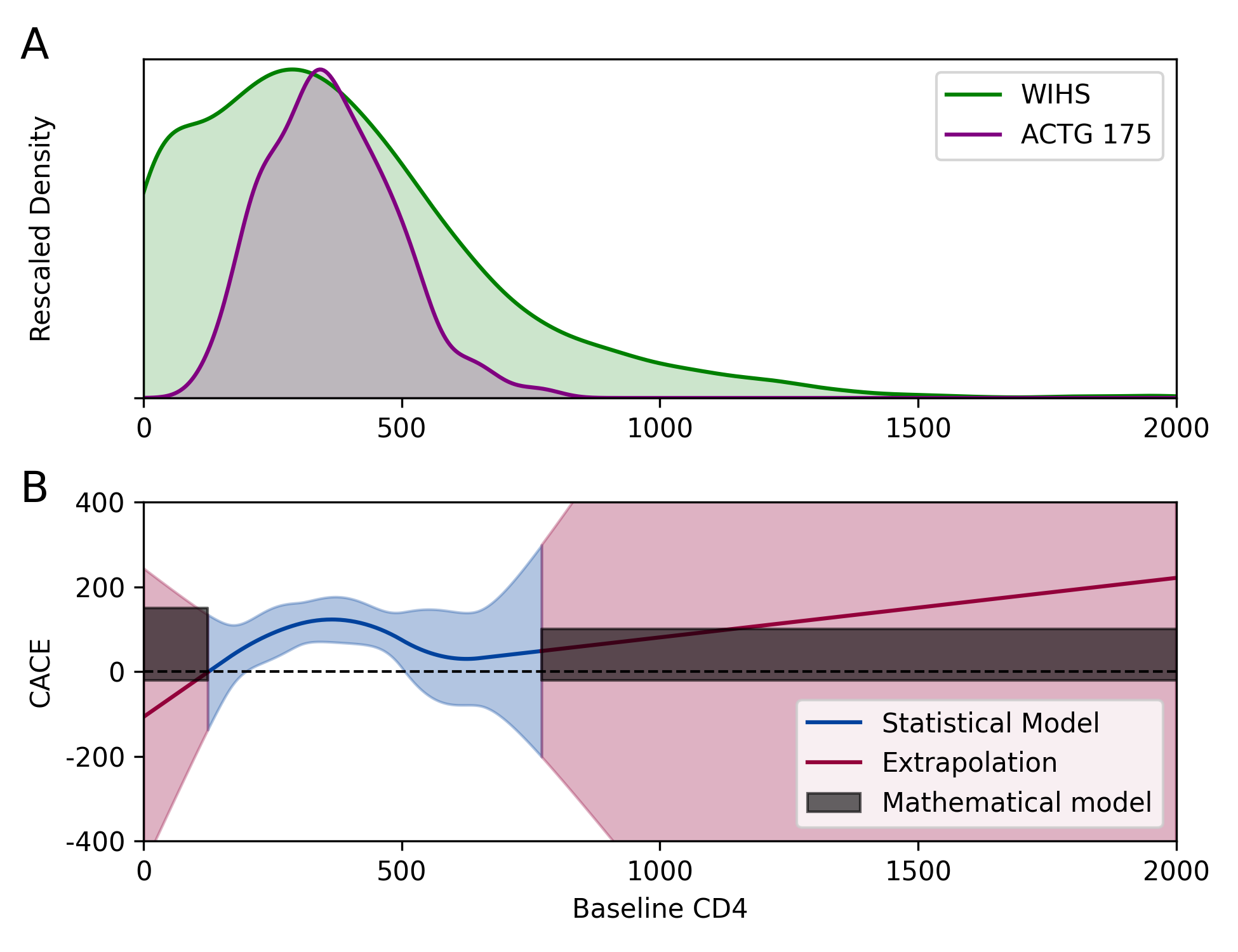}
	\floatfoot{ACTG: AIDS Clinical Trial Group, CACE: Conditional Average Causal Effect, WIHS: Women’s Interagency HIV Study. \\
		A: Baseline CD4 (cells/mm3) by data source. Densities were rescaled to have the same maximum height. \\
		B: The statistical model corresponds to the estimated CACE in the WIHS target population. The statistical model was estimated by restricting the WIHS data to the range of the observed baseline CD4 counts in ACTG 175. The extrapolation corresponds to extension of the CACE model to the full range of CD4 counts in WIHS. Note that this extrapolation differs from the applied extrapolation results. Shaded regions for the statistical model and extrapolation indicate 95\% confidence intervals estimated via the empirical sandwich variance estimator. The mathematical model corresponds to the range of CACE consistent with the chosen parameters for the nonpositive regions.
	}
	\label{Fig2}
\end{figure}

To harmonize the data, the ACTG 175 trial was restricted to women. There was a single participant in the ACTG 175 trial who had a baseline CD4 count of zero, with the next closest observation being 124. To prevent overfitting to this single observation, this participant was dropped from the analysis. In \cite{dahabreh_sensitivity_2023}, the WIHS cohort was restricted to participants with baseline CD4 cell counts between 200 and 500 to address nonpositivity. Instead, the approaches reviewed in Section 3 are demonstrated by making inference to the full WIHS target population. Differences in baseline CD4 were treated as structural positivity violations, despite the boundary points not being sharply defined due to the discrepancy between screening CD4 and baseline CD4 in the ACTG 175. Baseline CD4 in the ACTG 175 ranged from 124 to 771, whereas the range in WIHS was 0 to 1972 (Figure \ref{Fig2}A). Therefore, two nonpositive regions were present in this example.

\subsection{Statistical Analyses}

To address the motivating question with the synthesis approach and two nonpositive regions, notice that
\[\psi = \psi_0 \Pr(V<124 \mid R=1) + \psi_1 \Pr(124 \le V \le 771 \mid R=1) + \psi_2 \Pr(V > 771 \mid R=1)\]
where $V$ is baseline CD4, $\psi_0 = E[Y^1 - Y^0 \mid V < 124, R=1]$, $\psi_1 = E[Y^1 - Y^0 \mid 124 \le V \le 771, R=1]$, and $\psi_2 = E[Y^1 - Y^0 \mid V > 771, R=1]$. Then, $\psi_1$ was estimated using a statistical model and $\psi_0,\psi_2$ were replaced with mathematical models here. For the synthesis approach, the CACE estimator was selected due to the feasibility of specifying parameters for the mathematical model with external information contemporaneous to the data sources used. The overall synthesis model was specified as
\[\mathcal{G}(O; \gamma, \eta, \delta) = \delta_1 I(V < 124) + \mathpzc{s}(O; \gamma, \eta) I(124 \le V \le 771) + \delta_2 I(V > 771)\]
The statistical model component used the same model specification as the restricted target population and extrapolation approaches. The CACE model consisted of baseline CD4 modeled via restricted quadratic splines. Therefore, the mathematical model corresponds to the difference in 20-week CD4 when comparing two-drug versus one-drug ART for each of the nonpositive regions. For the mathematical model, bounds on the possible values for $\delta_1$ and $\delta_2$ were used (as opposed to distributions). Based on pharmacokinetic studies of two-drug therapies at the time \citep{collier_combination_1993,wilde_zidovudine_1993,meng_combination_1992}, one would not expect the considered two-drug ART combinations to result in lower CD4 cell counts at 20 weeks compared to one-drug ART. Therefore, one could suppose that the lowest the CACE would be in the nonpositive regions is zero. To be slightly conservative in our lower bound, we set $\delta_1 = \delta_2 = -20$. This choice of lower bound corresponds to a mild antagonistic interaction between drugs, such that the efficacy is reduced relative to zidovudine-only in the nonpositive regions. As an upper bound, $\delta_1 = 150$ was chosen based on the upper end of increases in CD4 observed in small scale studies over similar time frames for those with lower baseline CD4 counts \citep{collier_combination_1993}. For those with higher baseline CD4, the benefit of two-drug ART was assumed to be $\delta_2 = 100$, which was less stark but still provided a benefit over one-drug ART. How the choice of mathematical model was related to a statistical CACE model is shown in Figure \ref{Fig2}B. \edit{Since we were unwilling to differentiate between the possible values of $\delta$, the synthesis CACE estimator was used to construct bounds for $\psi$ instead of a point estimate. The estimated bounds for $\psi$ were computed by taking the infimum and supremum of the estimates from all unique combinations of $\delta$, as described in Section 3.1.3.}

For comparison, non-synthesis approaches were also applied. First, the mean difference in CD4 at 20 weeks between treatment groups from the ACTG 175 trial was used as an estimate of $\psi$. We refer to this as the na\"ive approach as it assumes women in ACTG 175 are a random sample of the same population as the WIHS cohort. Next, the restricted target population approach was applied by restricting the WIHS cohort to women with baseline CD4 counts between 124 and 771. For the restricted covariate set approach, baseline CD4 was excluded from both adjustment sets. Finally, we applied the extrapolation approach, which used the weighted regression AIPW outcome model to extrapolate. Age, weight, and baseline CD4 were all modeled using restricted quadratic splines with four knots for the sampling and outcome models. The outcome model further included interaction terms between baseline CD4 and assigned ART. 

\subsection{Results}

The ACTG 175 and WIHS participants differed in terms of age, baseline CD4, and race (Table \ref{Tab3}). As each of these variables was presupposed to be related to subsequent CD4 count, the ACTG 175 results were not expected to be immediately transportable to the target population. This concern regarding transportability was further supported by the differences in baseline CD4 (Figure \ref{Fig2}A) and how the CACE varied by baseline CD4 (Figure \ref{Fig2}B). However, most of the WIHS participants were in the positive region (71\%), with 20\% in the lower nonpositive region and 9\% in the upper nonpositive region. 

\begin{table}
	\caption{Descriptive statistics for the ACTG 175 trial and WIHS cohort}
	\centering
	\begin{tabular}{lcc} 
		\hline
		& ACTG 175 ($n_0 = 276$) & WIHS ($n_1 = 1932$)  \\ 
		\cline{2-3}
		Age (years)\textsuperscript{*}                    & 33 [28, 39]               & 36 [31,41]              \\
		Baseline CD4 count (cells/mm3)\textsuperscript{*} & 350 [278, 443]            & 330 [161, 516]          \\
		Weight (kilograms)\textsuperscript{*}             & 67 [59, 76]               & 66 [58, 78]             \\
		White                   				       & 154 (56\%)                & 390 (20\%)              \\
		Assigned two-drug ART          					& 175 (64\%)                & -                       \\
		CD4 at 20 weeks\textsuperscript{*}                & 357 [267, 480]            & -                       \\
		\hline
	\end{tabular}
	\floatfoot{ACTG: AIDS Clinical Trial Group, ART: antiretroviral therapy, WIHS: Women’s Interagency HIV Study. \\
	* Braces indicate the 25\textsuperscript{th} and 75\textsuperscript{th} percentiles
	}
	\label{Tab3}
\end{table}

\begin{figure}
	\centering
	\caption{Estimates of the average causal effect of two-drug versus one-drug antiretroviral therapy on CD4 count at 20 weeks by estimator}
	\includegraphics[width=0.90\linewidth]{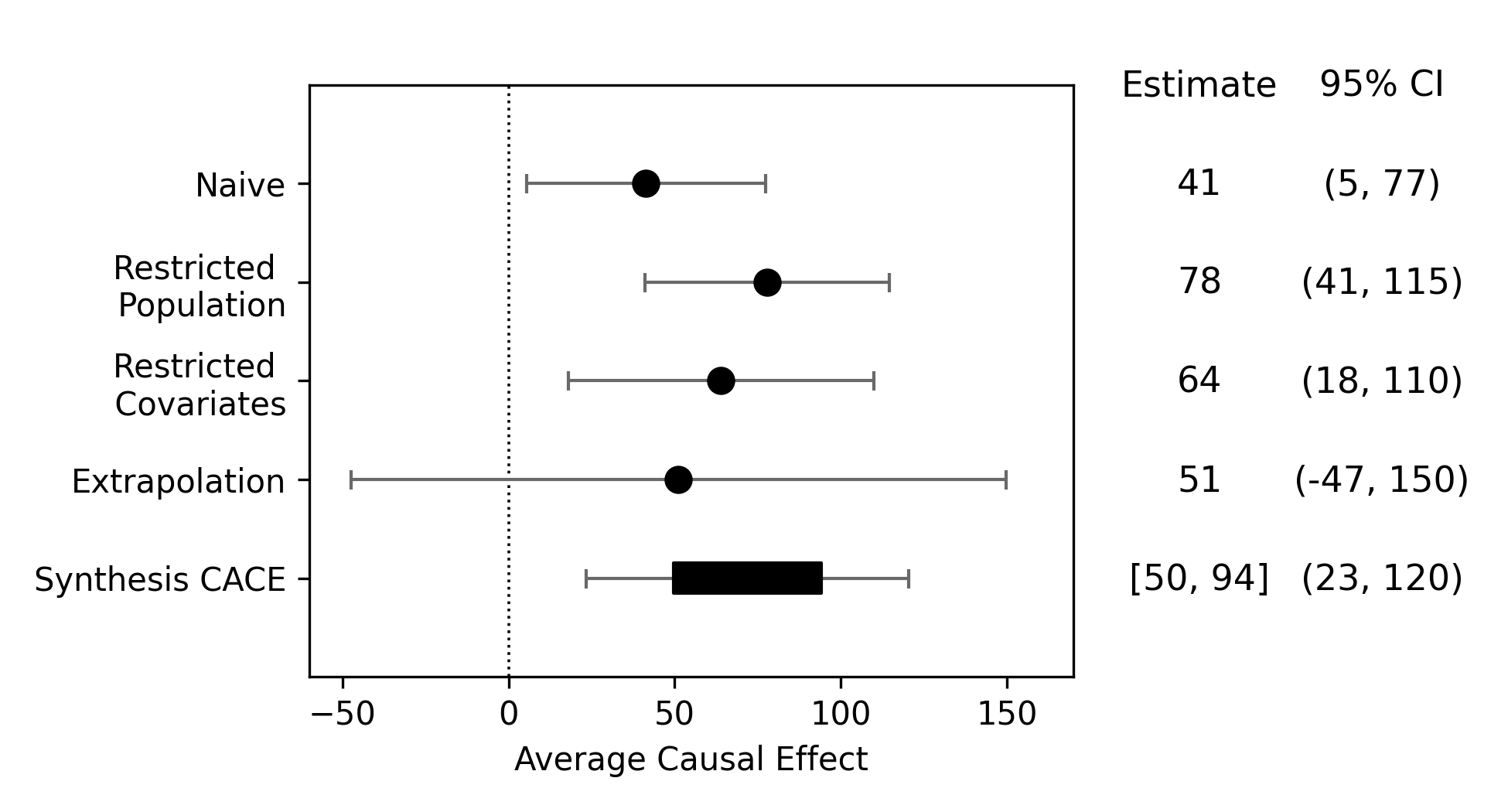}
	\floatfoot{CACE: condtional average causal effect, CI: confidence interval. Black circles indicate point estimates, black rectangles correspond to the bounds, and gray lines indicate the 95\% CI. }
	\label{Fig3}
\end{figure}

Results for the competing approaches are presented in Figure \ref{Fig3}. The bounds for the synthesis CACE estimator ranged from 50 to 94 cells/mm\textsuperscript{3}, with a lower CI of 23 and upper CI of 120. These results were consistent with a beneficial effect of two-drug over one-drug ART. The non-synthesis approaches produced varied results about how beneficial two-drug ART was over one-drug ART. The lowest point estimate was from the na\"ive method, which indicated an increase of 41 cells/mm\textsuperscript{3} (95\% CI: 5, 77) at 20 weeks. When restricting the WIHS target population to baseline CD4 counts between 124 and 771 (1375 observations remained in the WIHS data), a noticeably stronger effect was observed (78 cells/mm\textsuperscript{3}, 95\% CI: 41, 115). The restricted covariate set indicated an increase of 64 cells/mm\textsuperscript{3} (95\% CI: 18, 110). However, these results were suspect due to \edit{the likely occurrence of additive effect measure modification by} baseline CD4. The extrapolation approach resulted in a point estimate close to the na\"ive method (51 cells/mm\textsuperscript{3}), but results were highly uncertain (95\% CI: -47, 150). In fact, the extrapolation results were consistent with a harmful effect of two-drug ART on 20-week CD4 cell count versus one-drug ART. 

\section{Discussion}

Here, a statistical model was combined with a mathematical model to integrate multiple data sources in order to address structural positivity violations with a continuous covariate. Two novel AIPW estimators were introduced, assessed via simulation, and illustrated with an applied example from HIV. The proposed synthesis estimators had little bias and favorable coverage across the scenarios explored in the simulations, performing better than existing approaches to address nonpositivity (i.e., restricting the target population, restricting the covariate set, extrapolation). In the applied example, none of the non-synthesis methods were able to adequately address the motivating question without modifying the assumptions that lacked substantive justification. These results highlight the inadequacies of the non-synthesis approaches reviewed. When integrating contemporaneous external information, the synthesis CACE estimator indicates that two-drug ART would be beneficial on short-term CD4 cell counts in the target population, though this approach does not produce a point estimate without refinement of the mathematical model.

When considering synthesis methods in practice, it is crucial to recall the reliance on quality external information to fill in the nonpositive regions. \edit{Without this information, there is no justification for the choice of mathematical model. If invalid external information is used, estimates from synthesis models can be biased} \citep{zivich_transportability_2023} As also stated, potential outcomes were assumed to be well-defined for everyone in the target population. While this assumption can be reasonable (e.g., mortality under differing ART regimens for a target population of people with HIV that includes those with CD4 counts above a trial eligibility threshold), it may not always be (e.g., mortality under surgical removal of the prostate for a target population that includes persons without prostates). Historically, the distinction between nonpositivity due to unrealized possibilities versus logical impossibilities has not always been highlighted, but the synthesis framework is only capable of dealing with the former as there is no basis for building a mathematical model for the latter. What is deemed logically possible today may not always be, but it is important for the actions considered to be at least verifiable in principle (i.e., one should be able to imagine a study capable of addressing the question). When presenting results, it may be preferred to provide both the restricted target population (which can correspond to the statistical model portion of the synthesis estimator) and synthesis model. Presenting additional variations on the chosen parameters for the mathematical model and how results change is also likely to be beneficial.

While the core concepts of synthesizing statistical and mathematical models have existed \citep{wright_semi-empirical_nodate,rahmstorf_semi-empirical_2007,rezaei_hybrid_2020}, the motivation behind this approach to address positivity violations is relatively new \citep{zivich_transportability_2023}. This novelty means there are multiple avenues for future research. Two broad areas are reviewed here. First are theoretical extensions of the synthesis framework. Conditional exchangeability assumptions are often used to address other systematic errors (e.g., confounding, missing data, measurement error), and each of those exchangeability assumptions is paired with a positivity assumption. Using the synthesis framework to address violations of these other positivity assumptions is of interest. Other areas include accommodating different outcome types (e.g., time-to-event) and more complex causal structures (e.g., time-varying confounding). As previously stated, extrapolation is commonly used to address \textit{random} positivity violations. Other approaches to positivity violations have attempted to shore up extrapolations for random positivity through priors \citep{zhu_addressing_2023}. In a similar way, the synthesis framework provides an opportunity to address random nonpositivity.

The second general area is further development of synthesis estimators. Regarding the statistical model, g-computation, IPW, and AIPW estimators have been proposed here and in other work \citep{zivich_transportability_2023}. However, one could also presumably apply targeted maximum likelihood estimators \citep{schuler_targeted_2017}, or g-estimation of structural nested models \citep{naimi_introduction_2017}. For estimation of nuisance parameters, finite dimension parametric models were used. Use of machine learning algorithms for estimation of nuisance parameters in the statistical models is also of interest. Regarding the mathematical model, there are many opportunities for further study. This and previous work have focused on mathematical models that modify the statistical model. One could also consider using the mathematical model to regularize, or penalize, the statistical model in the positive regions to improve precision. Comparisons against other mathematical model structures that do not modify the statistical model (e.g., mechanistic models) is also an open area. As shown in Appendix 2, the statistical model component in the synthesis AIPW estimators is multiply robust. Whether and how mathematical models could be made robust to modeling assumptions is also important. \edit{There is also interest in studying how misspecification of the mathematical model impacts the overall bias for $\psi$.} Finally, diagnostic procedures and sensitivity analyses for both the statistical and mathematical models could help increase the reliability of synthesis estimators. One option would be to use the mathematical model in the positive region and compare to the observed data. While this approach does not indicate the validity of the mathematical model in the nonpositive regions, it may still serve as a useful check.

\section*{Acknowledgments}

This work was supported by the US National Institutes of Health under award numbers K01-AI177102 (PNZ), R01-AI157758 (PNZ, JKE, BES, SRC), R01-GM140564 (PNZ, JKE, JL), P30-AI050410 (PNZ, BES, SRC), and R35-GM147013 (ETL). The content is solely the responsibility of the authors and does not necessarily represent the official views of the National Institutes of Health. 

Computing code to replicate the illustrative example and the simulation experiment is available at https://github.com/pzivich/publications-code.
Data used for the illustrative example are available on GitHub (ACTG 175) and from https://statepi.jhsph.edu/wihs/wordpress/?page\_id=10771 (WIHS). 

\bibliographystyle{plainnat}
\bibliography{StatMechModel}{}

\begin{thebibliography}{74}
\providecommand{\natexlab}[1]{#1}
\providecommand{\url}[1]{\texttt{#1}}
\expandafter\ifx\csname urlstyle\endcsname\relax
  \providecommand{\doi}[1]{doi: #1}\else
  \providecommand{\doi}{doi: \begingroup \urlstyle{rm}\Url}\fi

\bibitem[Çağlayan et~al.(2018)Çağlayan, Terawaki, Chen, Rai, Ayer, and
  Flowers]{caglayan_microsimulation_2018}
Çağlar Çağlayan, Hiromi Terawaki, Qiushi Chen, Ashish Rai, Turgay Ayer, and
  Christopher~R. Flowers.
\newblock Microsimulation {Modeling} in {Oncology}.
\newblock \emph{JCO Clin Cancer Inform}, 2:\penalty0 CCI.17.00029, March 2018.

\bibitem[Bang and Robins(2005)]{bang_doubly_2005}
Heejung Bang and James~M Robins.
\newblock Doubly robust estimation in missing data and causal inference models.
\newblock \emph{Biometrics}, 61\penalty0 (4):\penalty0 962--973, 2005.
\newblock ISSN 0006-341X.

\bibitem[Bareinboim and Pearl(2016)]{bareinboim_causal_2016}
Elias Bareinboim and Judea Pearl.
\newblock Causal inference and the data-fusion problem.
\newblock \emph{Proceedings of the National Academy of Sciences}, 113\penalty0
  (27):\penalty0 7345--7352, 2016.
\newblock \doi{10.1073/pnas.1510507113}.

\bibitem[Barkan et~al.(1998)Barkan, Melnick, Preston-Martin, Weber, Kalish,
  Miotti, Young, Greenblatt, Sacks, and Feldman]{barkan_womens_1998}
S.~E. Barkan, S.~L. Melnick, S.~Preston-Martin, K.~Weber, L.~A. Kalish,
  P.~Miotti, M.~Young, R.~Greenblatt, H.~Sacks, and J.~Feldman.
\newblock The {Women}'s {Interagency} {HIV} {Study}. {WIHS} {Collaborative}
  {Study} {Group}.
\newblock \emph{Epidemiology}, 9\penalty0 (2):\penalty0 117--125, March 1998.

\bibitem[Bockting et~al.(2023)Bockting, Radev, and
  Bürkner]{bockting_simulation-based_2023}
Florence Bockting, Stefan~T. Radev, and Paul-Christian Bürkner.
\newblock Simulation-{Based} {Prior} {Knowledge} {Elicitation} for {Parametric}
  {Bayesian} {Models}, August 2023.

\bibitem[Cole and Frangakis(2009)]{cole_consistency_2009}
Stephen~R Cole and Constantine~E Frangakis.
\newblock The consistency statement in causal inference: a definition or an
  assumption?
\newblock \emph{Epidemiology}, 20\penalty0 (1):\penalty0 3--5, 2009.
\newblock ISSN 1044-3983.

\bibitem[Cole and Hernán(2008)]{cole_constructing_2008}
Stephen~R. Cole and Miguel~A. Hernán.
\newblock Constructing inverse probability weights for marginal structural
  models.
\newblock \emph{Am J Epidemiol}, 168\penalty0 (6):\penalty0 656--664, 2008.
\newblock ISSN 1476-6256 0002-9262.
\newblock \doi{10.1093/aje/kwn164}.
\newblock Edition: 08/05.

\bibitem[Cole et~al.(2019)Cole, Hudgens, Edwards, Brookhart, Richardson,
  Westreich, and Adimora]{cole_nonparametric_2019}
Stephen~R. Cole, Michael~G. Hudgens, Jessie~K. Edwards, M.~Alan Brookhart,
  David~B. Richardson, Daniel Westreich, and Adaora~A. Adimora.
\newblock Nonparametric {Bounds} for the {Risk} {Function}.
\newblock \emph{American Journal of Epidemiology}, 188\penalty0 (4):\penalty0
  632--636, 2019.
\newblock ISSN 0002-9262.
\newblock \doi{10.1093/aje/kwz013}.

\bibitem[Collier et~al.(1993)Collier, Coombs, Fischl, Skolnik, Northfelt,
  Boutin, Hooper, Kaplan, Volberding, Davis, Henrard, Weller, and
  Corey]{collier_combination_1993}
Ann~C. Collier, Robert~W. Coombs, Margaret~A. Fischl, Paul~R. Skolnik, Donald
  Northfelt, Paul Boutin, Carol~J. Hooper, Lawrence~D. Kaplan, Paul~A.
  Volberding, L.~Gray Davis, Denis~R. Henrard, Stephen Weller, and Lawrence
  Corey.
\newblock Combination {Therapy} with {Zidovudine} and {Didanosine} {Compared}
  with {Zidovudine} {Alone} in {HIV}-1 {Infection}.
\newblock \emph{Ann Intern Med}, 119\penalty0 (8):\penalty0 786--793, October
  1993.
\newblock ISSN 0003-4819.
\newblock Publisher: American College of Physicians.

\bibitem[Dahabreh et~al.(2019)Dahabreh, Robertson, Tchetgen~Tchetgen, Stuart,
  and Hernán]{dahabreh_generalizing_2019}
Issa~J. Dahabreh, Sarah~E. Robertson, Eric~J. Tchetgen~Tchetgen, Elizabeth~A.
  Stuart, and Miguel~A. Hernán.
\newblock Generalizing causal inferences from individuals in randomized trials
  to all trial-eligible individuals.
\newblock \emph{Biometrics}, 75\penalty0 (2):\penalty0 685--694, 2019.
\newblock ISSN 1541-0420.
\newblock \doi{10.1111/biom.13009}.

\bibitem[Dahabreh et~al.(2020)Dahabreh, Robertson, Steingrimsson, Stuart, and
  Hernán]{dahabreh_extending_2020}
Issa~J. Dahabreh, Sarah~E. Robertson, Jon~A. Steingrimsson, Elizabeth~A.
  Stuart, and Miguel~A. Hernán.
\newblock Extending inferences from a randomized trial to a new target
  population.
\newblock \emph{Statistics in Medicine}, 39\penalty0 (14):\penalty0 1999--2014,
  2020.
\newblock ISSN 1097-0258.
\newblock \doi{10.1002/sim.8426}.

\bibitem[Dahabreh et~al.(2023)Dahabreh, Robins, Haneuse, Saeed, Robertson,
  Stuart, and Hernán]{dahabreh_sensitivity_2023}
Issa~J. Dahabreh, James~M. Robins, Sebastien J.-P.~A. Haneuse, Iman Saeed,
  Sarah~E. Robertson, Elizabeth~A. Stuart, and Miguel~A. Hernán.
\newblock Sensitivity analysis using bias functions for studies extending
  inferences from a randomized trial to a target population.
\newblock \emph{Statistics in Medicine}, 42\penalty0 (13):\penalty0 2029--2043,
  2023.
\newblock ISSN 1097-0258.

\bibitem[Degtiar and Rose(2023)]{degtiar_review_2023}
Irina Degtiar and Sherri Rose.
\newblock A {Review} of {Generalizability} and {Transportability}.
\newblock \emph{Annual Review of Statistics and Its Application}, 10\penalty0
  (1), 2023.

\bibitem[Fox et~al.(2021)Fox, MacLehose, and Lash]{fox_introduction_2021}
Matthew~P. Fox, Richard~F. MacLehose, and Timothy~L. Lash.
\newblock Introduction, {Objectives}, and an {Alternative}.
\newblock In Matthew~P. Fox, Richard~F. MacLehose, and Timothy~L. Lash,
  editors, \emph{Applying {Quantitative} {Bias} {Analysis} to {Epidemiologic}
  {Data}}, Statistics for {Biology} and {Health}, pages 1--24. Springer
  International Publishing, Cham, 2021.

\bibitem[Fréchet(1935)]{frechet_generalisation_1935}
Maurice Fréchet.
\newblock Généralisation du théorème des probabilités totales.
\newblock \emph{Fundamenta Mathematicae}, 25\penalty0 (1):\penalty0 379--387,
  1935.
\newblock ISSN 0016-2736.

\bibitem[Gail et~al.(1984)Gail, Wieand, and Piantadosi]{gail_biased_1984}
M.~H. Gail, S.~Wieand, and S.~Piantadosi.
\newblock Biased estimates of treatment effect in randomized experiments with
  nonlinear regressions and omitted covariates.
\newblock \emph{Biometrika}, 71\penalty0 (3):\penalty0 431--444, December 1984.

\bibitem[Good(1992)]{good_bayesnon-bayes_1992}
I.~J. Good.
\newblock The {Bayes}/{Non}-{Bayes} {Compromise}: {A} {Brief} {Review}.
\newblock \emph{Journal of the American Statistical Association}, 87\penalty0
  (419):\penalty0 597--606, September 1992.
\newblock ISSN 0162-1459.
\newblock \doi{10.1080/01621459.1992.10475256}.

\bibitem[Greenland(2006)]{greenland_bayesian_2006}
Sander Greenland.
\newblock Bayesian perspectives for epidemiological research: {I}.
  {Foundations} and basic methods.
\newblock \emph{International Journal of Epidemiology}, 35\penalty0
  (3):\penalty0 765--775, June 2006.
\newblock ISSN 0300-5771.
\newblock \doi{10.1093/ije/dyi312}.

\bibitem[Greenland et~al.(1999)Greenland, Pearl, and
  Robins]{greenland_confounding_1999}
Sander Greenland, Judea Pearl, and James~M. Robins.
\newblock Confounding and {Collapsibility} in {Causal} {Inference}.
\newblock \emph{Statistical Science}, 14\penalty0 (1):\penalty0 29--46,
  February 1999.
\newblock ISSN 0883-4237, 2168-8745.
\newblock \doi{10.1214/ss/1009211805}.

\bibitem[Hammer et~al.(1996)Hammer, Katzenstein, Hughes, Gundacker, Schooley,
  Haubrich, Henry, Lederman, Phair, Niu, Hirsch, and
  Merigan]{hammer_trial_1996}
Scott~M. Hammer, David~A. Katzenstein, Michael~D. Hughes, Holly Gundacker,
  Robert~T. Schooley, Richard~H. Haubrich, W.~Keith Henry, Michael~M. Lederman,
  John~P. Phair, Manette Niu, Martin~S. Hirsch, and Thomas~C. Merigan.
\newblock A {Trial} {Comparing} {Nucleoside} {Monotherapy} with {Combination}
  {Therapy} in {HIV}-{Infected} {Adults} with {CD4} {Cell} {Counts} from 200 to
  500 per {Cubic} {Millimeter}.
\newblock \emph{New England Journal of Medicine}, 335\penalty0 (15):\penalty0
  1081--1090, October 1996.
\newblock ISSN 0028-4793.
\newblock \doi{10.1056/NEJM199610103351501}.

\bibitem[Hammer et~al.(1997)Hammer, Squires, Hughes, Grimes, Demeter, Currier,
  Eron, Feinberg, Balfour, Deyton, Chodakewitz, Fischl, Phair, Pedneault,
  Nguyen, and Cook]{hammer_controlled_1997}
Scott~M. Hammer, Kathleen~E. Squires, Michael~D. Hughes, Janet~M. Grimes,
  Lisa~M. Demeter, Judith~S. Currier, Joseph~J. Eron, Judith~E. Feinberg,
  Henry~H. Balfour, Lawrence~R. Deyton, Jeffrey~A. Chodakewitz, Margaret~A.
  Fischl, John~P. Phair, Louise Pedneault, Bach-Yen Nguyen, and Jon~C. Cook.
\newblock A {Controlled} {Trial} of {Two} {Nucleoside} {Analogues} plus
  {Indinavir} in {Persons} with {Human} {Immunodeficiency} {Virus} {Infection}
  and {CD4} {Cell} {Counts} of 200 per {Cubic} {Millimeter} or {Less}.
\newblock \emph{N Engl J Med}, 337\penalty0 (11):\penalty0 725--733, September
  1997.
\newblock ISSN 0028-4793.
\newblock \doi{10.1056/NEJM199709113371101}.
\newblock Publisher: Massachusetts Medical Society.

\bibitem[Harris et~al.(2020)Harris, Millman, van~der Walt, Gommers, Virtanen,
  Cournapeau, Wieser, Taylor, Berg, Smith, Kern, Picus, Hoyer, van Kerkwijk,
  Brett, Haldane, del Río, Wiebe, Peterson, Gérard-Marchant, Sheppard, Reddy,
  Weckesser, Abbasi, Gohlke, and Oliphant]{harris_array_2020}
Charles~R. Harris, K.~Jarrod Millman, Stéfan~J. van~der Walt, Ralf Gommers,
  Pauli Virtanen, David Cournapeau, Eric Wieser, Julian Taylor, Sebastian Berg,
  Nathaniel~J. Smith, Robert Kern, Matti Picus, Stephan Hoyer, Marten~H. van
  Kerkwijk, Matthew Brett, Allan Haldane, Jaime~Fernández del Río, Mark
  Wiebe, Pearu Peterson, Pierre Gérard-Marchant, Kevin Sheppard, Tyler Reddy,
  Warren Weckesser, Hameer Abbasi, Christoph Gohlke, and Travis~E. Oliphant.
\newblock Array programming with {NumPy}.
\newblock \emph{Nature}, 585\penalty0 (7825):\penalty0 357--362, September
  2020.
\newblock ISSN 1476-4687.
\newblock \doi{10.1038/s41586-020-2649-2}.

\bibitem[Hernán and Robins(2006)]{hernan_estimating_2006}
Miguel~A. Hernán and James~M. Robins.
\newblock Estimating causal effects from epidemiological data.
\newblock \emph{Journal of Epidemiology \& Community Health}, 60\penalty0
  (7):\penalty0 578--586, July 2006.
\newblock ISSN 0143-005X, 1470-2738.
\newblock \doi{10.1136/jech.2004.029496}.

\bibitem[Icazatti et~al.(2023)Icazatti, Abril-Pla, Klami, and
  Martin]{icazatti_preliz_2023}
Alejandro Icazatti, Oriol Abril-Pla, Arto Klami, and Osvaldo~A. Martin.
\newblock {PreliZ}: {A} tool-box for prior elicitation.
\newblock \emph{Journal of Open Source Software}, 8\penalty0 (89):\penalty0
  5499, September 2023.
\newblock ISSN 2475-9066.

\bibitem[Kennedy(2019)]{kennedy_nonparametric_2019}
Edward~H. Kennedy.
\newblock Nonparametric {Causal} {Effects} {Based} on {Incremental}
  {Propensity} {Score} {Interventions}.
\newblock \emph{Journal of the American Statistical Association}, 114\penalty0
  (526):\penalty0 645--656, April 2019.
\newblock ISSN 0162-1459.
\newblock \doi{10.1080/01621459.2017.1422737}.
\newblock Publisher: Taylor \& Francis \_eprint:
  https://doi.org/10.1080/01621459.2017.1422737.

\bibitem[Kennedy(2020)]{kennedy_optimal_2020}
Edward~H. Kennedy.
\newblock Optimal doubly robust estimation of heterogeneous causal effects.
\newblock \emph{arXiv:2004.14497 [math, stat]}, June 2020.

\bibitem[King and Zeng(2006)]{king2006dangers}
Gary King and Langche Zeng.
\newblock The {D}angers of {E}xtreme {C}ounterfactuals.
\newblock \emph{Political {A}nalysis}, 14\penalty0 (2):\penalty0 131--159,
  2006.

\bibitem[Kirkeby et~al.(2021)Kirkeby, Brookes, Ward, Dürr, and
  Halasa]{kirkeby_practical_2021}
Carsten Kirkeby, Victoria~J. Brookes, Michael~P. Ward, Salome Dürr, and Tariq
  Halasa.
\newblock A {Practical} {Introduction} to {Mechanistic} {Modeling} of {Disease}
  {Transmission} in {Veterinary} {Science}.
\newblock \emph{Front Vet Sci}, 7:\penalty0 546651, January 2021.

\bibitem[Krijkamp et~al.(2018)Krijkamp, Alarid-Escudero, Enns, Jalal, Hunink,
  and Pechlivanoglou]{krijkamp_microsimulation_2018}
Eline~M. Krijkamp, Fernando Alarid-Escudero, Eva~A. Enns, Hawre~J. Jalal,
  M.G.~Myriam Hunink, and Petros Pechlivanoglou.
\newblock Microsimulation modeling for health decision sciences using {R}: a
  tutorial.
\newblock \emph{Med Decis Making}, 38\penalty0 (3):\penalty0 400--422, April
  2018.
\newblock \doi{10.1177/0272989X18754513}.

\bibitem[Lesko et~al.(2016)Lesko, Cole, Hall, Westreich, Miller, Eron, Li,
  Mugavero, and {for the CNICS Investigators}]{lesko_effect_2016}
Catherine~R Lesko, Stephen~R Cole, H~Irene Hall, Daniel Westreich, William~C
  Miller, Joseph~J Eron, Jianmin Li, Michael~J Mugavero, and {for the CNICS
  Investigators}.
\newblock The effect of antiretroviral therapy on all-cause mortality,
  generalized to persons diagnosed with {HIV} in the {USA}, 2009–11.
\newblock \emph{International Journal of Epidemiology}, 45\penalty0
  (1):\penalty0 140--150, February 2016.

\bibitem[Lessler and Cummings(2016)]{lessler_mechanistic_2016}
Justin Lessler and Derek A.~T. Cummings.
\newblock Mechanistic {Models} of {Infectious} {Disease} and {Their} {Impact}
  on {Public} {Health}.
\newblock \emph{American Journal of Epidemiology}, 183\penalty0 (5):\penalty0
  415--422, March 2016.

\bibitem[Li et~al.(2019)Li, Thomas, and Li]{li_addressing_2019}
Fan Li, Laine~E. Thomas, and Fan Li.
\newblock Addressing {Extreme} {Propensity} {Scores} via the {Overlap}
  {Weights}.
\newblock \emph{Am J Epidemiol}, 188\penalty0 (1):\penalty0 250--257, January
  2019.
\newblock ISSN 1476-6256.
\newblock \doi{10.1093/aje/kwy201}.

\bibitem[Maldonado and Greenland(2002)]{maldonado_estimating_2002}
George Maldonado and Sander Greenland.
\newblock Estimating causal effects.
\newblock \emph{International Journal of Epidemiology}, 31\penalty0
  (2):\penalty0 422--429, April 2002.
\newblock ISSN 0300-5771.
\newblock \doi{10.1093/ije/31.2.422}.

\bibitem[Manski(1990)]{manski_nonparametric_1990}
Charles~F. Manski.
\newblock Nonparametric {Bounds} on {Treatment} {Effects}.
\newblock \emph{The American Economic Review}, 80\penalty0 (2):\penalty0
  319--323, 1990.
\newblock ISSN 0002-8282.
\newblock Publisher: American Economic Association.

\bibitem[McKinney(2010)]{mckinney_data_2010}
Wes McKinney.
\newblock Data {Structures} for {Statistical} {Computing} in {Python}.
\newblock \emph{Proceedings of the 9th Python in Science Conference}, pages
  56--61, 2010.
\newblock \doi{10.25080/Majora-92bf1922-00a}.
\newblock Conference Name: Proceedings of the 9th Python in Science Conference.

\bibitem[Meng et~al.(1992)Meng, Fischl, Boota, Spector, Bennett, Bassiakos,
  Lai, Wright, and Richman]{meng_combination_1992}
Tze-Chiang Meng, Margaret~A. Fischl, Ahmad~M. Boota, Stephen~A. Spector, Donald
  Bennett, Yiannis Bassiakos, Shenghan Lai, Brian Wright, and Douglas~D.
  Richman.
\newblock Combination {Therapy} with {Zidovudine} and {Dideoxycytidine} in
  {Patients} with {Advanced} {Human} {Immunodeficiency} {Virus} {Infection}.
\newblock \emph{Ann Intern Med}, 116\penalty0 (1):\penalty0 13--20, January
  1992.

\bibitem[Mikkola et~al.(2023)Mikkola, Martin, Chandramouli, Hartmann, Pla,
  Thomas, Pesonen, Corander, Vehtari, Kaski, Bürkner, and
  Klami]{mikkola_prior_2023}
Petrus Mikkola, Osvaldo~A. Martin, Suyog Chandramouli, Marcelo Hartmann,
  Oriol~Abril Pla, Owen Thomas, Henri Pesonen, Jukka Corander, Aki Vehtari,
  Samuel Kaski, Paul-Christian Bürkner, and Arto Klami.
\newblock Prior {Knowledge} {Elicitation}: {The} {Past}, {Present}, and
  {Future}.
\newblock \emph{Bayesian Analysis}, pages 1--33, January 2023.

\bibitem[Morris et~al.(2019)Morris, White, and Crowther]{morris_using_2019}
Tim~P. Morris, Ian~R. White, and Michael~J. Crowther.
\newblock Using simulation studies to evaluate statistical methods.
\newblock \emph{Statistics in Medicine}, 38\penalty0 (11):\penalty0 2074--2102,
  2019.
\newblock ISSN 1097-0258.
\newblock \doi{10.1002/sim.8086}.

\bibitem[Naimi et~al.(2017)Naimi, Cole, and Kennedy]{naimi_introduction_2017}
Ashley~I. Naimi, Stephen~R. Cole, and Edward~H. Kennedy.
\newblock An introduction to g methods.
\newblock \emph{Int J Epidemiol}, 46\penalty0 (2):\penalty0 756--762, 2017.
\newblock ISSN 1464-3685 0300-5771.
\newblock \doi{10.1093/ije/dyw323}.

\bibitem[Naimi et~al.(2021)Naimi, Rudolph, Kennedy, Cartus, Kirkpatrick, Haas,
  Simhan, and Bodnar]{naimi_incremental_2021}
Ashley~I. Naimi, Jacqueline~E. Rudolph, Edward~H. Kennedy, Abigail Cartus,
  Sharon~I. Kirkpatrick, David~M. Haas, Hyagriv Simhan, and Lisa~M. Bodnar.
\newblock Incremental {Propensity} {Score} {Effects} for {Time}-fixed
  {Exposures}.
\newblock \emph{Epidemiology}, 32\penalty0 (2):\penalty0 202--208, March 2021.
\newblock ISSN 1044-3983.
\newblock \doi{10.1097/EDE.0000000000001315}.

\bibitem[Nethery et~al.(2019)Nethery, Mealli, and
  Dominici]{nethery_estimating_2019}
Rachel~C. Nethery, Fabrizia Mealli, and Francesca Dominici.
\newblock Estimating {P}opulation {A}verage {C}ausal {E}ffects in the
  {P}resence of {N}on-{O}verlap: The {E}ffect of {N}atural {G}as {C}ompressor
  {S}tation {E}xposure on {C}ancer {M}ortality.
\newblock \emph{Ann Appl Stat}, 13\penalty0 (2):\penalty0 1242--1267, June
  2019.
\newblock ISSN 1932-6157.

\bibitem[Nguyen et~al.(2017)Nguyen, Ebnesajjad, Cole, and
  Stuart]{nguyen_sensitivity_2017}
Trang~Quynh Nguyen, Cyrus Ebnesajjad, Stephen~R. Cole, and Elizabeth~A. Stuart.
\newblock Sensitivity {Analysis} for an {Unobserved} {Moderator} in
  {Rct}-to-{Target}-{Population} {Generalization} of {Treatment} {Effects}.
\newblock \emph{The Annals of Applied Statistics}, 11\penalty0 (1):\penalty0
  225--247, 2017.
\newblock ISSN 1932-6157.

\bibitem[Nguyen et~al.(2018)Nguyen, Ackerman, Schmid, Cole, and
  Stuart]{nguyen_sensitivity_2018}
Trang~Quynh Nguyen, Benjamin Ackerman, Ian Schmid, Stephen~R. Cole, and
  Elizabeth~A. Stuart.
\newblock Sensitivity analyses for effect modifiers not observed in the target
  population when generalizing treatment effects from a randomized controlled
  trial: {Assumptions}, models, effect scales, data scenarios, and
  implementation details.
\newblock \emph{PLOS ONE}, 13\penalty0 (12):\penalty0 e0208795, December 2018.
\newblock ISSN 1932-6203.

\bibitem[Nilsson et~al.(2023)Nilsson, Björk, and Bonander]{nilsson_proxy_2023}
Anton Nilsson, Jonas Björk, and Carl Bonander.
\newblock Proxy {Variables} and the {Generalizability} of {Study} {Results}.
\newblock \emph{American Journal of Epidemiology}, 192\penalty0 (3):\penalty0
  448--454, February 2023.

\bibitem[O’Hagan(2019)]{ohagan_expert_2019}
Anthony O’Hagan.
\newblock Expert {Knowledge} {Elicitation}: {Subjective} but {Scientific}.
\newblock \emph{The American Statistician}, 73\penalty0 (sup1):\penalty0
  69--81, March 2019.

\bibitem[Petersen et~al.(2012)Petersen, Porter, Gruber, Wang, and van~der
  Laan]{petersen_diagnosing_2012}
Maya~L Petersen, Kristin~E Porter, Susan Gruber, Yue Wang, and Mark~J van~der
  Laan.
\newblock Diagnosing and responding to violations in the positivity assumption.
\newblock \emph{Stat Methods Med Res}, 21\penalty0 (1):\penalty0 31--54, 2012.
\newblock ISSN 0962-2802.
\newblock \doi{10.1177/0962280210386207}.

\bibitem[Rahmstorf(2007)]{rahmstorf_semi-empirical_2007}
Stefan Rahmstorf.
\newblock A {Semi}-{Empirical} {Approach} to {Projecting} {Future}
  {Sea}-{Level} {Rise}.
\newblock \emph{Science}, 315\penalty0 (5810):\penalty0 368--370, January 2007.

\bibitem[Railsback and Grimm(2011)]{railsback_agent-based_2011}
Steven~F. Railsback and Volker Grimm.
\newblock \emph{Agent-{Based} and {Individual}-{Based} {Modeling}: {A}
  {Practical} {Introduction}}.
\newblock Princeton University Press, 2011.

\bibitem[Rezaei et~al.(2020)Rezaei, Hayduk, Alkan, Kemski, Delebinski, and
  Bertram]{rezaei_hybrid_2020}
Reza Rezaei, Christopher Hayduk, Emre Alkan, Thomas Kemski, Thaddaeus
  Delebinski, and Christoph Bertram.
\newblock Hybrid {Phenomenological} and {Mathematical}-{Based} {Modeling}
  {Approach} for {Diesel} {Emission} {Prediction}.
\newblock In \emph{{SAE} {Technical} {Paper}}, April 2020.

\bibitem[Roberts et~al.(2012)Roberts, Russell, Paltiel, Chambers, McEwan, and
  Krahn]{roberts_conceptualizing_2012}
Mark Roberts, Louise~B. Russell, A.~David Paltiel, Michael Chambers, Phil
  McEwan, and Murray Krahn.
\newblock Conceptualizing a {Model}: {A} {Report} of the {ISPOR}-{SMDM}
  {Modeling} {Good} {Research} {Practices} {Task} {Force}–2.
\newblock \emph{Med Decis Making}, 32\penalty0 (5):\penalty0 678--689,
  September 2012.

\bibitem[Robins et~al.(2000)Robins, Hernan, and Brumback]{robins_marginal_2000}
J.~M. Robins, M.~A. Hernan, and B.~Brumback.
\newblock Marginal structural models and causal inference in epidemiology.
\newblock \emph{Epidemiology (Cambridge, Mass.)}, 11\penalty0 (5):\penalty0
  550--60, September 2000.
\newblock ISSN 1044-3983 (Print) 1044-3983.
\newblock Edition: 2000/08/24.

\bibitem[Robins et~al.(2007)Robins, Sued, Lei-Gomez, and
  Rotnitzky]{robins_comment_2007}
James Robins, Mariela Sued, Quanhong Lei-Gomez, and Andrea Rotnitzky.
\newblock Comment: {Performance} of {Double}-{Robust} {Estimators} {When}
  “{Inverse} {Probability}” {Weights} {Are} {Highly} {Variable}.
\newblock \emph{Statistical Science}, 22\penalty0 (4):\penalty0 544--559,
  November 2007.
\newblock ISSN 0883-4237, 2168-8745.
\newblock \doi{10.1214/07-STS227D}.
\newblock Publisher: Institute of Mathematical Statistics.

\bibitem[Schuler and Rose(2017)]{schuler_targeted_2017}
Megan~S Schuler and Sherri Rose.
\newblock Targeted maximum likelihood estimation for causal inference in
  observational studies.
\newblock \emph{Am J Epidemiol}, 185\penalty0 (1):\penalty0 65--73, 2017.
\newblock ISSN 0002-9262.

\bibitem[Shepherd et~al.(2008)Shepherd, Redman, and
  Ankerst]{shepherd_does_2008}
Bryan~E. Shepherd, Mary~W. Redman, and Donna~P. Ankerst.
\newblock Does {Finasteride} {Affect} the {Severity} of {Prostate} {Cancer}?
  {A} {Causal} {Sensitivity} {Analysis}.
\newblock \emph{Journal of the American Statistical Association}, 103\penalty0
  (484):\penalty0 1392--1404, December 2008.
\newblock ISSN 0162-1459.
\newblock \doi{10.1198/016214508000000706}.

\bibitem[Slayton et~al.(2020)Slayton, O’Hagan, Barnes, Rhea, Hilscher, Rubin,
  Lofgren, Singh, Segre, Paul, and {for the Centers for Disease Control and
  Prevention MInD-Healthcare Program}]{slayton_modeling_2020}
Rachel~B Slayton, Justin~J O’Hagan, Sean Barnes, Sarah Rhea, Rainer Hilscher,
  Michael Rubin, Eric Lofgren, Brajendra Singh, Alberto Segre, Prabasaj Paul,
  and {for the Centers for Disease Control and Prevention MInD-Healthcare
  Program}.
\newblock Modeling {Infectious} {Diseases} in {Healthcare} {Network}
  ({MInD}-{Healthcare}) {Framework} for {Describing} and {Reporting}
  {Multidrug}-resistant {Organism} and {Healthcare}-{Associated} {Infections}
  {Agent}-based {Modeling} {Methods}.
\newblock \emph{Clinical Infectious Diseases}, 71\penalty0 (9):\penalty0
  2527--2532, November 2020.

\bibitem[Snowden et~al.(2011)Snowden, Rose, and
  Mortimer]{snowden_implementation_2011}
Jonathan~M. Snowden, Sherri Rose, and Kathleen~M. Mortimer.
\newblock Implementation of {G}-computation on a simulated data set:
  demonstration of a causal inference technique.
\newblock \emph{Am J Epidemiol}, 173\penalty0 (7):\penalty0 731--738, 2011.
\newblock ISSN 1476-6256 0002-9262.
\newblock \doi{10.1093/aje/kwq472}.
\newblock Edition: 03/16.

\bibitem[Stefanski and Boos(2002)]{stefanski_calculus_2002}
Leonard~A Stefanski and Dennis~D Boos.
\newblock The {Calculus} of {M}-{Estimation}.
\newblock \emph{The American Statistician}, 56\penalty0 (1):\penalty0 29--38,
  February 2002.
\newblock ISSN 0003-1305.
\newblock \doi{10.1198/000313002753631330}.

\bibitem[Stürmer et~al.(2010)Stürmer, Rothman, Avorn, and
  Glynn]{sturmer_treatment_2010}
Til Stürmer, Kenneth~J. Rothman, Jerry Avorn, and Robert~J. Glynn.
\newblock Treatment {Effects} in the {Presence} of {Unmeasured} {Confounding}:
  {Dealing} {With} {Observations} in the {Tails} of the {Propensity} {Score}
  {Distribution}—{A} {Simulation} {Study}.
\newblock \emph{Am J Epidemiol}, 172\penalty0 (7):\penalty0 843--854, October
  2010.
\newblock ISSN 0002-9262.

\bibitem[Tchetgen~Tchetgen et~al.(2020)Tchetgen~Tchetgen, Ying, Cui, Shi, and
  Miao]{tchetgen_tchetgen_introduction_2020}
Eric~J. Tchetgen~Tchetgen, Andrew Ying, Yifan Cui, Xu~Shi, and Wang Miao.
\newblock An {Introduction} to {Proximal} {Causal} {Learning}.
\newblock \emph{arXiv:2009.10982 [stat]}, September 2020.
\newblock arXiv: 2009.10982.

\bibitem[van~der Laan and Luedtke(2015)]{van_der_laan_targeted_2015}
Mark~J. van~der Laan and Alexander~R. Luedtke.
\newblock Targeted {Learning} of the {Mean} {Outcome} under an {Optimal}
  {Dynamic} {Treatment} {Rule}.
\newblock \emph{J Causal Inference}, 3\penalty0 (1):\penalty0 61--95, March
  2015.
\newblock ISSN 2193-3677.
\newblock \doi{10.1515/jci-2013-0022}.

\bibitem[Van~der Laan and Robins(2003)]{vdL2003}
Mark~J Van~der Laan and James~M Robins.
\newblock \emph{Unified methods for censored longitudinal data and causality},
  volume~5.
\newblock Springer, 2003.

\bibitem[Vansteelandt and Keiding(2011)]{vansteelandt_invited_2011}
Stijn Vansteelandt and Niels Keiding.
\newblock Invited {Commentary}: {G}-{Computation}–{Lost} in {Translation}?
\newblock \emph{American Journal of Epidemiology}, 173\penalty0 (7):\penalty0
  739--742, April 2011.
\newblock ISSN 0002-9262.
\newblock \doi{10.1093/aje/kwq474}.

\bibitem[Vansteelandt et~al.(2006)Vansteelandt, Goetghebeur, Kenward, and
  Molenberghs]{vansteelandt_ignorance_2006}
Stijn Vansteelandt, Els Goetghebeur, Michael~G. Kenward, and Geert Molenberghs.
\newblock Ignorance and {Uncertainty} {Regions} as {Inferential} {Tools} in a
  {Sensitivity} {Analysis}.
\newblock \emph{Statistica Sinica}, 16\penalty0 (3):\penalty0 953--979, 2006.
\newblock ISSN 1017-0405.

\bibitem[Virtanen et~al.(2020)Virtanen, Gommers, Oliphant, Haberland, Reddy,
  Cournapeau, Burovski, Peterson, Weckesser, and Bright]{virtanen_scipy_2020}
Pauli Virtanen, Ralf Gommers, Travis~E Oliphant, Matt Haberland, Tyler Reddy,
  David Cournapeau, Evgeni Burovski, Pearu Peterson, Warren Weckesser, and
  Jonathan Bright.
\newblock {SciPy} 1.0: fundamental algorithms for scientific computing in
  {Python}.
\newblock \emph{Nature Methods}, 17\penalty0 (3):\penalty0 261--272, 2020.
\newblock ISSN 1548-7105.

\bibitem[Webster-Clark and Breskin(2021)]{webster-clark_directed_2021}
Michael Webster-Clark and Alexander Breskin.
\newblock Directed {Acyclic} {Graphs}, {Effect} {Measure} {Modification}, and
  {Generalizability}.
\newblock \emph{Am J Epidemiol}, 190\penalty0 (2):\penalty0 322--327, February
  2021.
\newblock ISSN 1476-6256.
\newblock \doi{10.1093/aje/kwaa185}.

\bibitem[Westreich and Cole(2010)]{westreich_invited_2010}
Daniel Westreich and Stephen~R Cole.
\newblock Invited commentary: positivity in practice.
\newblock \emph{American Journal of Epidemiology}, 171\penalty0 (6):\penalty0
  674--677, 2010.
\newblock ISSN 1476-6256.

\bibitem[Westreich et~al.(2017)Westreich, Edwards, Lesko, Stuart, and
  Cole]{westreich_transportability_2017}
Daniel Westreich, Jessie~K Edwards, Catherine~R Lesko, Elizabeth Stuart, and
  Stephen~R Cole.
\newblock Transportability of {Trial} {Results} {Using} {Inverse} {Odds} of
  {Sampling} {Weights}.
\newblock \emph{Am J Epidemiol}, 186\penalty0 (8):\penalty0 1010--1014, October
  2017.
\newblock ISSN 0002-9262.
\newblock \doi{10.1093/aje/kwx164}.

\bibitem[Wilde and Langtry(1993)]{wilde_zidovudine_1993}
Michelle~I. Wilde and Heather~D. Langtry.
\newblock Zidovudine.
\newblock \emph{Drugs}, 46\penalty0 (3):\penalty0 515--578, September 1993.
\newblock ISSN 1179-1950.

\bibitem[Wright and Potapczuk(2004)]{wright_semi-empirical_nodate}
William Wright and Mark Potapczuk.
\newblock Semi-{Empirical} {Modelling} of {SLD} {Physics}.
\newblock In \emph{42nd {AIAA} {Aerospace} {Sciences} {Meeting} and {Exhibit}}.
  American Institute of Aeronautics and Astronautics, 2004.
\newblock \doi{10.2514/6.2004-412}.

\bibitem[Zhu et~al.(2023)Zhu, Mitra, and Roy]{zhu_addressing_2023}
Angela~Yaqian Zhu, Nandita Mitra, and Jason Roy.
\newblock Addressing positivity violations in causal effect estimation using
  {Gaussian} process priors.
\newblock \emph{Statistics in Medicine}, 42\penalty0 (1):\penalty0 33--51,
  2023.

\bibitem[Zhu et~al.(2021)Zhu, Hubbard, Chubak, Roy, and Mitra]{zhu_core_2021}
Yaqian Zhu, Rebecca~A. Hubbard, Jessica Chubak, Jason Roy, and Nandita Mitra.
\newblock Core {Concepts} in {Pharmacoepidemiology}: {Violations} of the
  {Positivity} {Assumption} in the {Causal} {Analysis} of {Observational}
  {Data}: {Consequences} and {Statistical} {Approaches}.
\newblock \emph{Pharmacoepidemiol Drug Saf}, 30\penalty0 (11):\penalty0
  1471--1485, November 2021.

\bibitem[Zivich et~al.(2022{\natexlab{a}})Zivich, Cole, and
  Westreich]{zivich_positivity_2022}
Paul~N. Zivich, Stephen~R. Cole, and Daniel Westreich.
\newblock Positivity: {Identifiability} and {Estimability}.
\newblock \emph{arXiv:2203.11300 [stat]}, July 2022{\natexlab{a}}.
\newblock \doi{10.48550/arXiv.2207.05010}.

\bibitem[Zivich et~al.(2022{\natexlab{b}})Zivich, Klose, Cole, Edwards, and
  Shook-Sa]{zivich_delicatessen_2022}
Paul~N. Zivich, Mark Klose, Stephen~R. Cole, Jessie~K. Edwards, and Bonnie~E.
  Shook-Sa.
\newblock Delicatessen: {M}-{Estimation} in {Python}.
\newblock \emph{arXiv:2203.11300 [stat]}, March 2022{\natexlab{b}}.

\bibitem[Zivich et~al.(2024)Zivich, Edwards, Lofgren, Cole, Shook-Sa, and
  Lessler]{zivich_transportability_2023}
Paul~N. Zivich, Jessie~K. Edwards, Eric~T. Lofgren, Stephen~R. Cole, Bonnie~E.
  Shook-Sa, and Justin Lessler.
\newblock Transportability without positivity: a synthesis of statistical and
  simulation modeling.
\newblock \emph{Epidemiology}, 35:\penalty0 23--31, January 2024.
\newblock \doi{10.1097/EDE.0000000000001677}.

\end{thebibliography}

\newpage

\section*{Appendix}

\subsection*{Appendix 1: Identification of $\psi_0$}

Here, the identification result given in Section 3.1 is proven. For $a \in \{0,1\}$,
\begin{equation*}
	\begin{split}
		E[Y^a \mid V^*=0, R=1] = & E[E[Y^a \mid V,W,V^*=0, R=1] | V^*=0,R=1] \\
		= & E[E[Y^a \mid V,W,V^*=0, R=0] | V^*=0,R=1] \\
		= & E[E[Y^a \mid V,W, R=0] | V^*=0,R=1] \\
		= & E[E[Y^a \mid A=a,V,W, R=0] | V^*=0,R=1] \\
		= & E[E[Y \mid A=a,V,W, R=0] | V^*=0,R=1] \\
	\end{split}
\end{equation*}
which follows from iterated expectations, (\ref{Eq4a}) and (\ref{Eq5a}), $V^* = R$ when $V^*=0$, (\ref{Eq2}) and (\ref{Eq3}), and (\ref{Eq1}). Notice that (\ref{Eq2}) and (\ref{Eq3}) do not need to be modified due to those assumptions already being restricted to values of $v$ observed in the external population.

\subsection*{Appendix 2: Consistency and Asymptotic Normality of the Synthesis Estimators}

Here, consistency and asymptotic normality of the two proposed synthesis estimators is demonstrated. Before considering the estimators, note the following relation when $B$ is a binary variable,
\begin{equation}
	E[AB \mid C] = E[A \mid C,B=1] \Pr(B=1 \mid C)
	\tag{A1}
	\label{EqA2}
\end{equation}
This relationship is used throughout the following proofs.

\subsection*{Nuisance Model Estimators}
To demonstrate consistency and asymptotic normality of the proposed synthesis AIPW estimators, we first demonstrate that the estimating equations are unbiased under the stated assumptions. Since the statistical component of both synthesis estimators consist of estimating equations for the nuisance parameters of the weighted regression AIPW estimator, these estimating equations are considered first. The nuisance models for the weighted regression AIPW estimator can be expressed as the following set of stacked estimating equations,
\begin{equation}
	\phi(O_i; \theta) = 
	\begin{bmatrix}
		\phi_{\eta_1} \\ 
		\phi_{\eta_2} \\ 
		\phi_{\eta_3} \\ 
	\end{bmatrix}
	=
	\begin{bmatrix}
		(1 - R_i) \left\{ A_i - \pi_A(V_i,W_i,\eta_1) \right\} \mathbb{Z}_i^T \\
		(1 - V_i^*) \left\{ R_i - \pi_R(V_i,W_i,\eta_2) \right\}\mathbb{U}_i^T \\	
		\pi(A_i,V_i,W_i; \eta_1, \eta_2) \left\{ Y_i - q(A,V,W; \eta_3) \right\} \mathbb{X}_i^T \\	
	\end{bmatrix}
	\tag{A2}
	\label{EqA1}
\end{equation}
where
\[\pi(A_i,V_i,W_i; \eta_1, \eta_2) = \frac{\pi_R(V_i,W_i; \eta_2)}{1 - \pi_R(V_i,W_i; \eta_2)} \times I(R_i=0) \times \left[ \frac{A_i}{\pi_A(V_i,W_i;\eta_1)} + \frac{1-A_i}{1 - \pi_A(V_i,W_i;\eta_1)} \right] \]
\[\pi_A(V_i,W_i,\eta_1) = \text{expit}\left( \mathbb{Z}_i \eta_1^T \right)\]
\[\pi_R(V_i,W_i,\eta_2) = \text{expit}\left( \mathbb{U}_i \eta_2^T \right)\]
\[q(A_i,V_i,W_i;\eta_3) = \mathbb{X}_i \eta_3^T\]
Following maximum likelihood theory, $\phi_a$ and $\phi_r$ are unbiased with solutions $\hat{\eta}_1$ and $\hat{\eta}_2$, respectively. For ease of later notation, we shorten $\pi(A_i,V_i,W_i; \eta_1, \eta_2)$ to $\pi$, $\pi_A(V_i,W_i; \eta_1)$ to $\pi_A$, and $\pi_R(V_i,W_i; \eta_2)$ to $\pi_R$ hereafter. Further, we drop the subscripts on random variables. 

For $\phi_{\eta_3}$, first consider when the outcome model is correctly specified without any assumptions regarding the specification of the weight model. Here, the estimating equation for a generic element of $\mathbb{X}_i$, denoted by $\mathbb{x}_i$, is considered. Following application of the relation in (\ref{EqA2}), $E[\phi_{\eta_3}]$ becomes
\begin{equation}
	E\left[ \frac{\pi_R}{1-\pi_R} \times \left[ \frac{A}{\pi_A} + \frac{1-A}{1-\pi_A} \right] \left\{ Y - q(A,V,W;\eta_3) \right\}\mathbb{x}^T \mid R=0 \right] \Pr(R=0)
	\tag{A3}
	\label{EqA3}
\end{equation}
Following iterated expectation by $V,W$,
\begin{equation}
	E\left[ \frac{\pi_R}{1-\pi_R} \; E\left\{
	\left[ \frac{A}{\pi_A} + \frac{1-A}{1-\pi_A} \right] \left\{ Y - q(A,V,W;\eta_3) \right\}\mathbb{x}^T \mid V,W,R=0
	\right\} \mid R=0 \right] \Pr(R=0)
	\tag{A4}
	\label{EqA4}
\end{equation}
Consider the inner expectation for ease of presentation. Since $\pi_A$ is conditional on $V,W,R=0$, it follows that the inner expectation of (\ref{EqA4}) is equal to
\begin{equation}
	\begin{aligned}
		& \frac{1}{\pi_A}
		E\left[ A \left\{ Y - q(A,V,W;\eta_3) \right\}\mathbb{x}^T \mid V,W,R=0 \right] \\
		& +	\frac{1}{1-\pi_A}
		E\left[ (1-A) \left\{ Y - q(A,V,W;\eta_3) \right\}\mathbb{x}^T \mid V,W,R=0 \right]
	\end{aligned}
	\tag{A5}
	\label{EqA5}
\end{equation}
and following (\ref{EqA2}), (\ref{EqA5}) is equal to
\begin{equation}
	\begin{aligned}
		& \frac{\Pr(A=1 | V,W,R=0)}{\pi_A} \mathbb{x}^T
		E\left[ Y - q(A=1,V,W;\eta_3) \mid A=1,V,W,R=0 \right] \\
		& +	\frac{\Pr(A=0 | V,W,R=0)}{1-\pi_A}
		\mathbb{x}^T E\left[ Y - q(A=0,V,W;\eta_3) \mid A=0,V,W,R=0 \right]
	\end{aligned}
	\tag{A6}
	\label{EqA6}
\end{equation}
Focusing on the expectation, $E\left[ Y - q(A=a,V,W;\eta_3) \mid A=a,V,W,R=0 \right]$ for $a \in \{0,1\}$, it follows that
\begin{equation}
	\begin{aligned}
		E\left[ Y  - q(A=a,V,W;\eta_3) \mid A=a,V,W,R=0 \right] = & E\left[ Y \mid A=a,V,W,R=0 \right] - q(A=a,V,W;\eta_3) \\
		= & 0
	\end{aligned}
	\tag{A7}
	\label{EqA7}
\end{equation}
given conditional independence and correct outcome model specification, respectively. Plugging (\ref{EqA7}) into (\ref{EqA6}) and (\ref{EqA4}), it becomes clear that the estimating equation is unbiased regardless of the specification of $\pi_R$ and $\pi_A$ for all $\mathbb{x}$ in $\mathbb{X}$. As shown, $E\left[ Y \mid A=a,V,W,R=0 \right] = E\left[ Y^a \mid V,W,R=1,V^*=0 \right]$ under the corresponding identification assumptions.

Now consider the case when the weight models are both correctly specified but the outcome model may not be. For the following, assume that $\mathbb{X}$ includes at least an intercept term and main effect term for $A$, as stated in \cite{robins_comment_2007}. Otherwise, no restrictions are made on the outcome model. For ease of presentation, the estimating equation corresponding to the intercept in $\mathbb{X}$ is considered (i.e., $\mathbb{x} = 1$) without a loss of generality. Under these assumptions, the proof of unbiasedness for $E[\phi_{\eta_3}]$ proceeds using the same steps as above through (\ref{EqA6}). Under the assumption that the action model is correctly specified, i.e., $\pi_A = \Pr(A=1 | V,W,R=0)$, (\ref{EqA6}) instead reduces to
\begin{equation}
	E\left[ Y - q(A=1,V,W;\eta_3) \mid A=1,V,W,R=0 \right]
	+ E\left[ Y - q(A=0,V,W;\eta_3) \mid A=0,V,W,R=0 \right]
	\tag{A8}
	\label{EqA8}
\end{equation}
which further reduces to 
\begin{equation}
	E\left[ Y^1 \mid V,W,R=0 \right] - q(A=1,V,W;\eta_3) 
	+ E\left[ Y^0 \mid V,W,R=0 \right] - q(A=0,V,W;\eta_3)
	\tag{A9}
	\label{EqA9}
\end{equation}
following conditional independence, causal consistency (\ref{Eq1}), and action exchangeability with positivity (\ref{Eq2}, \ref{Eq3}). Plugging (\ref{EqA9}) into (\ref{EqA4}), results in
\begin{equation}
	\begin{aligned}
		& E\left[ \frac{\pi_R}{1-\pi_R} \; \left\{E\left[ Y^1 \mid V,W,R=0 \right] - q(A=1,V,W;\eta_3) \right\} \mid R=0 \right] \Pr(R=0) \\ 
		+ & E\left[ \frac{\pi_R}{1-\pi_R} \; \left\{ E\left[ Y^0 \mid V,W,R=0 \right] - q(A=0,V,W;\eta_3) \right\} \mid R=0 \right] \Pr(R=0)
	\end{aligned}
	\tag{A10}
	\label{EqA10}
\end{equation}
Now consider the parts of (\ref{EqA10}) for $a \in \{0,1\}$,
\begin{equation}
	E\left[ \frac{\pi_R}{1-\pi_R} \left\{ E\left[ Y^a \mid V,W,R=0 \right] - q(A=a,V,W;\eta_3) \right\} \mid R=0 \right] \Pr(R=0)
	\tag{A11}
	\label{EqA11}
\end{equation}
Equation (\ref{EqA11}) is equal to
\begin{equation}
	E\left[ \frac{\pi_R}{1-\pi_R} E\left[ Y^a \mid V,W,R=0 \right] - \frac{\pi_R}{1-\pi_R} q(A=a,V,W;\eta_3) \mid R=0 \right] \Pr(R=0)
	\tag{A12}
	\label{EqA12}
\end{equation}
\begin{equation}
	= E\left[ \frac{\pi_R}{1-\pi_R} E\left[ Y^a \mid V,W,R=0 \right] \mid R=0 \right] \Pr(R=0)
	- 
	E\left[ \frac{\pi_R (1-R)}{1-\pi_R} q(A=a,V,W;\eta_3)\right]
	\tag{A13}
	\label{EqA13}
\end{equation}
\begin{equation}
	= E\left[ E\left[ \frac{\pi_R}{1-\pi_R} Y^a \mid V,W,R=0 \right] \mid R=0 \right] \Pr(R=0)
	- 
	E\left[ \frac{\pi_R (1-R)}{1-\pi_R} q(A=a,V,W;\eta_3)\right]
	\tag{A14}
	\label{EqA14}
\end{equation}
\begin{equation}
	= E\left[ \frac{\pi_R}{1-\pi_R} Y^a \mid R=0 \right] \Pr(R=0)
	- 
	E\left[ \frac{\pi_R (1-R)}{1-\pi_R} q(A=a,V,W;\eta_3)\right]
	\tag{A15}
	\label{EqA15}
\end{equation}
\begin{equation}
	= E\left[ \frac{\pi_R(1-R)}{1-\pi_R} Y^a \right]
	- 
	E\left[ \frac{\pi_R (1-R)}{1-\pi_R} q(A=a,V,W;\eta_3)\right]
	\tag{A16}
	\label{EqA16}
\end{equation}
\begin{equation}
	= E\left[ \frac{\pi_R (1-R)}{1-\pi_R} \left\{Y^a - q(A=a,V,W;\eta_3) \right\} \right]
	\tag{A17}
	\label{EqA17}
\end{equation}
where (\ref{EqA13}) follows from summation of expectations and (\ref{EqA2}), (\ref{EqA14}) follows from conditional independence, (\ref{EqA15}) follows from undoing the iterative expectation, and (\ref{EqA16}) follows from (\ref{EqA2}). Now by iterative expectation by $V,W$ and conditional independence, (\ref{EqA17}) is equal to
\begin{equation}
	E\left[ \frac{\pi_R}{1-\pi_R} E \left\{(1-R)\left[Y^a - q(A=a,V,W;\eta_3)\right] \mid V,W \right\} \right]
	\tag{A18}
	\label{EqA18}
\end{equation}
and by (\ref{EqA2}), is equal to
\begin{equation}
	E\left[ \frac{\pi_R}{1-\pi_R} E \left\{Y^a - q(A=a,V,W;\eta_3) \mid V,W,R=0,V^*=0 \right\} \Pr(R=0 \mid V,W,V^*=0) \right]
	\tag{A19}
	\label{EqA19}
\end{equation}
Following correct model specification for the sampling model (i.e., $\pi_R = \Pr(R=1 | V,W,V^*=0)$), it follows that (\ref{EqA19}) is equal to
\begin{equation}
	E\left[ \Pr(R=1 \mid V,W,V^*=0) E\left\{Y^a - q(A=a,V,W;\eta_3) \mid V,W,R=0,V^*=0 \right\} \right]
	\tag{A20}
	\label{EqA20}
\end{equation}
Since $\Pr(R = 1 | V,W,V^*=0) >0$ by sampling positivity (\ref{Eq5a}), attention is limited to the innermost expectation of (\ref{EqA20}). Note that
\begin{equation}
	\begin{aligned}
		E\left\{Y^a - q(A=a,V,W;\eta_3) \mid V,W,R=0,V^*=0 \right\} = & E\left\{Y^a \mid V,W,R=0,V^*=0 \right\} - q(A=a,V,W;\eta_3) \\
		= & E\left\{Y^a \mid V,W,R=1,V^*=0 \right\} - q(A=a,V,W;\eta_3)
	\end{aligned}
	\tag{A21}
	\label{EqA21}
\end{equation}
by conditional independence, and sampling exchangeability with positivity (\ref{Eq4a}, \ref{Eq5a}). So, the solution for $E[\phi_{\eta_3}]$, $\hat{\eta}_3$, requires that $E\left\{Y^a \mid V,W,R=1,V^*=0 \right\} = q(A=a,V,W;\eta_3)$ for $a \in \{0,1\}$ regardless of the specification of the outcome model when both the weight models are correctly specified. Therefore, the third estimating equation is unbiased if either the outcome model is correctly specified or both weight models are correctly specified.

\subsection*{Synthesis MSM Estimator}

The synthesis MSM estimator consists of the estimating functions provided in (\ref{SynMSM}). As shown previously, $\phi_{\eta_1}$, $\phi_{\eta_2}$, and $\phi_{\eta_3}$ are unbiased. It follows via (\ref{EqA2}) that $E[\phi_{\alpha}]$ is equal to
\begin{equation}
	\begin{aligned}
		& E\left[ \left( \hat{Y}_i^1 - \mathbb{W}_i(1) \hat{\alpha}^T \right) \mathbb{w}_i(1)^T \mid R=1, V^*=0 \right] \Pr(R=1, V^*=0) \\
		+ & E\left[ \left( \hat{Y}_i^0 - \mathbb{W}_i(0) \hat{\alpha}^T \right) \mathbb{w}_i(0)^T \mid R=1, V^*=0 \right] \Pr(R=1, V^*=0)
	\end{aligned}
	\tag{A23}
	\label{EqA23}
\end{equation}
For ease of presentation, the intercept term of $\mathbb{W}(a)$ is considered (i.e., $\mathbb{w}(a) = 1$) without a loss of generality. Note that for $a \in \{0,1\}$, both expectation pieces of (\ref{EqA23}) are equal to
\begin{equation}
	\begin{aligned}
		E\left[ \left( \hat{Y}^a - \mathbb{W}(a) \alpha^T \right) \mid R=1, V^*=0 \right] = &
		E\left[ E\{Y^a | V,W,R=1,V^*=0\} - \mathbb{W}(a) \alpha^T \mid R=1, V^*=0 \right] \\
		= & E\left[ Y^a \mid R=1, V^*=0 \right] - E\left[ \mathbb{W}(a) \alpha^T \mid R=1, V^*=0 \right] \\
		= & E\left[ E\left\{ Y^a \mid V,R=1, V^*=0 \right\} \mid R=1, V^*=0 \right] - E\left[ \mathbb{W}(a)\alpha^T \mid R=1, V^*=0 \right] \\ 
		= & 0
	\end{aligned}
	\tag{A24}
	\label{EqA24}
\end{equation}
which follows from definition of $\hat{Y}^a$, summation of expectations and undoing the iterative expectation of $Y^a$, iterative expectation by $V$, and by definition of the MSM under correct model specification.

For $E[\phi_{\mu_s^1}]$, it follows by (\ref{EqA1}) that
\begin{equation}
	E[\mathbb{W}_i(1) \alpha^T + \mathbb{W}^*_i(1) \nu^T - \mu_s^1 \mid R=1] \Pr(R=1)
	\tag{A25}
	\label{EqA25}
\end{equation}
and by definition of $\mu_s^1$,
\begin{equation}
	E[\mathbb{W}_i(1) \alpha^T + \mathbb{W}^*_i(1) \nu^T \mid R=1] \Pr(R=1) -  E[Y^1 | R=1] \Pr(R=1)
	\tag{A26}
	\label{EqA26}
\end{equation}
Following the law of total probability, (\ref{EqA26}) is equal to
\begin{equation}
	\begin{aligned}
		& E[\mathbb{W}_i(1) \alpha^T + \mathbb{W}^*_i(1) \nu^T \mid R=1,V^*=0] \Pr(V^*=0 \mid R=1) \\
		+ & E[\mathbb{W}_i(1) \alpha^T + \mathbb{W}^*_i(1) \nu^T \mid R=1,V^*=1] \Pr(V^*=1 \mid R=1) \\
		- & E[Y^1 \mid R=1] \Pr(R=1)
	\end{aligned}
	\tag{A27}
	\label{EqA27}
\end{equation}
Now consider the first expectation in (\ref{EqA26}), 
\begin{equation}
	\begin{aligned}
		E[\mathbb{W}_i(1) \alpha^T + \mathbb{W}^*_i(1) \nu^T \mid R=1,V^*=0] = & E[\mathbb{W}_i(1) \alpha^T \mid R=1,V^*=0] \\
		= & E[E\{Y^1 \mid V,R=1,V^*=0\} \mid R=1,V^*=0] \\
		= & E[Y^1 \mid R=1,V^*=0]
	\end{aligned}	
	\tag{A28}
	\label{EqA28}
\end{equation}
follows via the mathematical model not contributing to the positive region (i.e., $\mathbb{W}_i^*(1) \nu^T = 0$ for $V^* = 0$), definition of $\mathbb{W}_i(1) \alpha^T$ under correct specification of the MSM, and undoing the iterative expectation. For the second expectation in (\ref{EqA27}), 
\begin{equation}
	\begin{aligned}
		E[\mathbb{W}_i(1) \alpha^T + \mathbb{W}_i(1) \nu^T \mid R=1,V^*=1] = & E[E\{Y^1 \mid V,R=1,V^*=1\} \mid R=1,V^*=1] \\
		= & E[Y^1 \mid R=1,V^*=1]
	\end{aligned}	
	\tag{A29}
	\label{EqA29}
\end{equation}
where the first equality follows from correct specification of the statistical MSM and that the mathematical model is correctly specified with the correct value of $\nu$, and the second follows from undoing the iterative expectation. Plugging the results of (\ref{EqA28}) and (\ref{EqA29}) into (\ref{EqA27}) and undoing the law of total expectation result in 
\begin{equation}
	\begin{aligned}
		\left( E[Y^1 \mid R=1] - E[Y^1 \mid R=1] \right) \Pr(R=1) = 0
	\end{aligned}	
	\tag{A30}
	\label{EqA30}
\end{equation}
A similar argument follows for $E[\phi_{\mu_s^0}]$. Finally, $\phi_{\psi_{sm}}$ is equal to zero and $E[\phi_{\psi_{sm}}]$ is unbiased. 

Since $\hat{\theta}$ is the solution to a vector of unbiased estimating equations, it follows under suitable regularity conditions that $\hat{\theta}$ is a consistent and asymptotically normal estimator of $\theta$ \cite{stefanski_calculus_2002} when paired with the correct value of $\nu$. Incorporating uncertainty into $\nu$ can be accomplished via the procedures described in the paper.

\subsection*{Synthesis CACE Estimator}

The synthesis CACE estimator consists of the estimating functions in (\ref{SynCACE}). Again, it follows from the prior proofs that $\phi_{\eta_1}$, $\phi_{\eta_2}$, and $\phi_{\eta_3}$ are unbiased under the corresponding assumptions. For the CACE model, $E[\phi_{\gamma}]$ is equal to
\begin{equation}
	E\left[ \left(\hat{Y}^1 - \hat{Y}^0\right) - \mathbb{V} \gamma^T \mid R=1, V^*=0 \right]
	\tag{A31}
	\label{EqA31}
\end{equation}
following (\ref{EqA2}). By definition of $\hat{Y}^a$ and correct specification of $\mathbb{V} \gamma^T$, (\ref{EqA31}) is equal to
\begin{equation}
	E\left[ E\left\{Y^1 - Y^0 \mid V,W,R=1,V^*=0 \right\} - E\left\{Y^1 - Y^0 \mid V,R=1,V^*=0 \right\} \mid R=1, V^*=0 \right] = 0
	\tag{A32}
	\label{EqA32}
\end{equation}
and thus is unbiased.

Following (\ref{EqA2}) and the definition of $\psi_{sc}$, $E[\phi_{\psi_{sc}}]$ is equal to
\begin{equation}
	E\left[ \mathbb{V} \gamma^T + \mathbb{V}^* \delta^T \mid R=1 \right]\Pr(R=1) - E[Y^1 - Y^0 \mid R=1] \Pr(R=1)
	\tag{A33}
	\label{EqA33}
\end{equation}
and following the law of total expectation is equal to
\begin{equation}
	\begin{aligned}
		& E\left[ \mathbb{V} \gamma^T + \mathbb{V}^* \delta^T \mid R=1,V^*=0 \right] \Pr(V^* = 0 \mid R=1) \\ 
		+ & E\left[ \mathbb{V} \gamma^T + \mathbb{V}^* \delta^T \mid R=1,V^*=1 \right] \Pr(V^* = 1 \mid R=1) \\
		- & E[Y^1 - Y^0 \mid R=1] \Pr(R=1)
	\end{aligned}
	\tag{A34}
	\label{EqA34}
\end{equation}
For the first expectation, it follows that
\begin{equation}
	\begin{aligned}
		E\left[ \mathbb{V} \gamma^T + \mathbb{V}^* \delta^T \mid R=1,V^*=0 \right]  = & E\left[ \mathbb{V} \gamma^T \mid R=1,V^*=0 \right] \\
		= & E\left[ E\{Y^1 - Y^0 \mid V,R=1,V^*=0\} \mid R=1,V^*=0 \right] \\ 
		= & E[Y^1 - Y^0 \mid R=1,V^*=0]
	\end{aligned}
	\tag{A35}
	\label{EqA35}
\end{equation}
by $\mathbb{V}_i^*(1) \nu^T = 0$ for $V^* = 0$, definition of $\mathbb{V} \gamma^T$ under correct model specification for the CACE model, and undoing the iterative expectation. For the second expectation, it similarly follows that 
\begin{equation}
	\begin{aligned}
		E\left[ \mathbb{V} \gamma^T + \mathbb{V}^* \delta^T \mid R=1,V^*=1 \right] = & E\left[ E\{Y^1 - Y^0 \mid V,R=1,V^*=1\} \mid R=1,V^*=1 \right] \\ 
		= & E[Y^1 - Y^0 \mid R=1,V^*=1]
	\end{aligned}
	\tag{A36}
	\label{EqA36}
\end{equation}
by correct specification of the statistical CACE model and that the mathematical model is correctly specified with the correct value of $\delta$, and undoing the iterative expectation. Plugging the results of (\ref{EqA35}) and (\ref{EqA36}) into (\ref{EqA34}) results in
\begin{equation}
	\left\{E[Y^1 - Y^0 \mid R=1] - E[Y^1 - Y^0 \mid R=1]\right\} \Pr(R=1) = 0
	\tag{A37}
	\label{EqA37}
\end{equation}
Therefore, $\hat{\theta}$ is the solution to a vector of unbiased estimating equations and the synthesis CACE estimator is consistent and asymptotically normal when paired with the correct value for $\delta$. Again, uncertainty in $\delta$ can be incorporated using the procedures reviewed.

\end{document}